\documentclass[12pt]{article}
\usepackage{graphicx}
\usepackage{bm}
\usepackage{amsmath,amssymb}
\usepackage{natbib}
\usepackage{subfigure}
\usepackage{url}
\usepackage{pifont}
\usepackage{booktabs}
\usepackage{verbatim}
\usepackage{ulem}
\usepackage{rotating}
\usepackage[margin=0.75 in]{geometry} 				
\usepackage{amsmath}
\usepackage{epsfig}
\usepackage{rotating}

\newcommand{\ben}{\begin{eqnarray}}
\newcommand{\een}{\end{eqnarray}}

\def\boxit#1{\vbox{\hrule\hbox{\vrule\kern6pt
            \vbox{\kern6pt#1\kern6pt}\kern6pt\vrule}\hrule}}

\def\bse{\begin{eqnarray*}}
\def\ese{\end{eqnarray*}}
\def\be{\begin{eqnarray}}
\def\ee{\end{eqnarray}}
\def\bq{\begin{equation}}
\def\eq{\end{equation}}
\def\bse{\begin{eqnarray*}}
\def\ese{\end{eqnarray*}}

\def\be{\begin{eqnarray}}
\def\ee{\end{eqnarray}}
\def\bq{\begin{equation}}
\def\eq{\end{equation}}
\def\bse{\begin{eqnarray*}}
\def\ese{\end{eqnarray*}}

\def\boxit#1{\vbox{\hrule\hbox{\vrule\kern6pt
            \vbox{\kern6pt#1\kern6pt}\kern6pt\vrule}\hrule}}
\usepackage{times}
\usepackage{epsfig}
\usepackage{rotating}

\usepackage{fancyhdr} 				
\usepackage{sectsty}
\allsectionsfont{\sffamily\mdseries\upshape} 

\usepackage[pdftex,dvipsnames,usenames]{color}

\def\boxit#1{\vbox{\hrule\hbox{\vrule\kern6pt\vbox{\kern6pt#1\kern6pt}\kern6pt\vrule}\hrule}}

\begin{document}

\baselineskip=14pt

\begin{center}
{\Large{ \bf      Bayesian sparse graphical models and their mixtures using lasso selection priors }}
\end{center}
\baselineskip=12pt

\begin{center}

Rajesh Talluri, Veerabhadran Baladandayuthapani  and  Bani K. Mallick. 
\let\thefootnote\relax\footnotetext{
Rajesh Talluri, is graduate student and Bani K. Mallick is Professor, Department of Statistics, Texas A\&M University, College Station, Texas, 77840. Veerabhadran Baladandayuthapani is Associate Professor Department of
Biostatistics, The University of Texas M.D. Anderson Cancer Center,Houston, Texas 77030.}
\vspace{2mm}

\end{center}


\begin{center}
{\Large{\bf Abstract}}
\end{center}
\baselineskip=12pt

We propose Bayesian methods for Gaussian graphical models that lead to sparse and adaptively shrunk estimators of the precision (inverse covariance) matrix.  Our methods are based on lasso-type regularization priors leading to parsimonious parameterization of the precision  matrix, which is essential in several applications involving learning relationships among the variables. In this context, we introduce a novel type of selection prior that develops a sparse structure on the precision matrix by making most of the elements exactly zero, in addition to ensuring positive definiteness -- thus conducting model selection and estimation simultaneously. We extend these methods to finite and infinite mixtures of Gaussian graphical models for clustered data using Dirichlet process priors. We discuss appropriate posterior simulation schemes to implement posterior inference in the proposed models, including the evaluation of normalizing constants that are functions of parameters of interest which result from the restrictions on the correlation matrix. We evaluate the operating characteristics of our method via several simulations and in application to real data sets.



\thispagestyle{empty}


\pagenumbering{arabic}
\newlength{\gnat}
\setlength{\gnat}{20pt}
\baselineskip=\gnat


\section{Introduction}
Consider the $p$ dimensional random vector ${\bm Y}=(Y^{(1)},\cdots ,Y^{(p)})$, which follows a multivariate normal distribution $N_{p}(\bm\mu,\bm\Sigma)$ where both the mean $\bm\mu$ and the variance-covariance matrix $\bm\Sigma$ are unknown.
Flexible modelling of the covariance matrix, $\bm\Sigma$, or equivalently the precision matrix, $\bm\Omega=\bm\Sigma^{-1}$,  is one of the most important tasks in analyzing Gaussian multivariate data. Furthermore, it has a direct relationship to constructing Gaussian graphical models (GGMs)  by identifying the significant edges. Of particular interest in this structure is the identification of zero entries in the precision matrix $\bm\Omega$. An off-diagonal zero entry $\Omega_{ij}=0$ indicates  conditional independence between the two random variables $Y^{(i)}$ and $Y^{(j)}$, given all other variables. This is the covariance selection problem or the model selection problem in the Gaussian graphical models \citep{dempster72,speed86,wong2003,yuan2007}, which provides a framework for the exploration of multivariate dependence patterns. GGMs are tools for modelling conditional independence relationships. Among the practical advantages of using GGMs in high-dimensional problems are their ability to  (i) make computations more efficient by alleviating the need to handle large matrices, (ii) yield better predictions by fitting sparser models, and (iii) aid scientific understanding by breaking down a global model into a collection of local models that are easier to search. Estimating the precision matrix efficiently and understanding its graphical structure is challenging, however, due to a variety of reasons that we discuss hereafter.

A GGM for a random vector $\bm{Y}$ can be represented by an undirected graph ${G} = (\bm{V} ,\bm{E})$, where $\bm{V}$ contains $p$ vertices corresponding to the $p$ variates and the edges $\bm{E} = (e_{ij} )_{(1\leq i< j \leq p)}$ describe the conditional independence relationships among $Y^{(1)},\ldots,Y^{(p)}$. The edge between $Y^{(i)}$ and $Y^{(j)}$ is absent if and only if $Y^{(i)}$ and $Y^{(j)}$ are independent, conditional on the other variables, and corresponds to $\Omega_{ij} = 0$. Thus, parameter estimation and model selection in the Gaussian  graphical model are equivalent to estimating parameters and identifying zeros in the precision matrix. The two main difficulties are that the number of unknown elements in the covariance matrix increases quadratically with $p$, and that it is difficult to deal directly with individual elements of the covariance matrix because it is necessary to keep the estimated matrix positive definite. \citet{berger94} and \citet{dempster69}  pointed out that estimators based on scalar multiples of the sample covariance matrix tend to distort the eigenstructure of the true covariance matrix unless $p/n$ is small. In this paper, we address these modelling and inferential challenges as we explore methods to adaptively estimate the precision matrix in a Gaussian graphical model setting. 

There have been many approaches to Gaussian graphical modelling. In a Bayesian setting, modelling  is based on hierarchical specifications for the covariance matrix  (or precision matrix) using  global conjugate priors on the space of positive-definite matrices, such as inverse Wishart priors or its equivalents.   \citet{lauritzen93} introduced an equivalent form as the hyper-inverse Wishart (HIW) distribution. Although that construction enjoys many advantages, such as computational efficiency due to its conjugate formulation and exact calculation of marginal likelihoods \citep{scott08}, it  is sometimes inflexible due to its restrictive form. Unrestricted graphical model determination is challenging unless the search space is restricted to decomposable graphs, where the marginal likelihoods are available up to the overall normalizing constants \citep{giudici96,roverato2000}. The marginal likelihoods are used to calculate the posterior probability of each graph, which gives an exact solution for small examples, but a prohibitively large number of graphs for a moderately large $p$. Moreover, extension to a nondecomposable graph is nontrivial and computationally expensive using reversible-jump algorithms \citep{giudici1999,Brooks03}. G-Wishart prior distributions has been proposed as an generalization of HIW priors  for nondecomposable graphs \citep{roverato2002, atay05}. Although it is convenient to use HIW prior due to its conjugate nature, it has several limitations. First, due to global nature of the hyper-prior specifications, HIW priors become more restrictive. Second, sampling from the non-decomposable HIW distribution is challenging due to the presence of an unknown normalizing constant. Computationally challenging  Monte Carlo based techniques have been proposed \citep{roverato2002,dellaportas2003, jones05, carvalho09, lenko11}. Due to this computational burden, extension of these models in a more complex framework (like mixture of graphical models)  will be a daunting task.

Alternate approaches for more adaptive estimation and/or selection in graphical models are based on methods that enforce sparsity either via variable (edge) selection  or regularization/penalization approaches. In a Bayesian regression context for variable selection problems such priors have been proposed by \citet{george93,george97,kuo98,della00,della02}. However the context of covariance selection in graphical models is inherently a different problem with additional complexity arising due to the additional constraints of positive definiteness and the number of parameters to estimate being in the the order of $p^2$ instead of $p$.  Regularization based covariance estimation has been done in a frequentist framework \citep{rothman2009, peng2009,levina2008}.  An alternate class of penalties that have received considerable attention in recent times have been lasso-type penalties \citet{tibshirani96} that have the ability to  promote sparseness, and have been used for variable selection in regression problems.  In a graphical model context, in a frequentist setting \citet{mein06}, \citet{yuan2007} and \citet{friedman08} proposed methods to estimate the precision or covariance matrix based on lasso-type penalties that yield only point estimates of the precision matrix. Lasso-based penalties are equivalent to Laplace priors in a Bayesian setting \citep{fig03,baemallick04,park05}. However, in a Bayesian setting, lasso penalties do not produce absolute zeros as the estimates of the precision matrix, and thus cannot be used to conduct model selection simultaneously in such settings. However \citet{wang2012} used Bayesian Lasso using a threshold to determine the zeros in the precision matrix and proposed an efficient Gibbs sampling scheme to estimated the precision matrix.

There have been several attempts to shrink the covariance/precision matrix via matrix factorizations for  unrestricted search over the space of both decomposable and nondecomposable graphs. \cite{barnard2000} factorized the covariance matrix in terms of standard deviations and correlations, proposed  several shrinkage estimators and discussed suitable priors. \citet{wong2003} expressed the inverse covariance matrix as a product of the inverse partial variances and the matrix of partial correlations, then used reversible-jump-based Markov chain Monte Carlo (MCMC) algorithms to identify the zeros among the diagonal elements. \citet{Liechty03} proposed flexible modelling schemes using decompositions of the correlation matrix

In this paper, we propose  novel Bayesian methods for GGMs that allow for simultaneous model selection and parameter estimation.  We introduce a novel type of  prior in Section 2 that can be decomposed into selection and shrinkage components in which lasso-type priors are used to accomplish shrinkage  and variable selection priors are used for selection.  We allow for local exploration of graphical dependencies that leads to a sparse structure of the precision matrix by enforcing most of the non-required elements to be exactly zero with positive probability while ensuring the estimate of the precision matrix is positive definite. More importantly, as a significant methodological innovation, we extend these methods to mixtures of GGMs for clustered data, with each mixture component assumed to be Gaussian with an adaptive covariance structure. Rodriguez et.al (2011)  developed similar methods for clustering data using dirichlet processes and Hidden Markov models, however our approach is different because we use shrinkage and selection on the precision matrices to cluster the data. For some kinds of data, it is reasonable to assume that the variables can be clustered or grouped based on sharing similar connectivity or graphs.  Our motivation for this model arises from a high-throughput gene expression data set, for which it is of interest not only to cluster the patients (samples) into the correct subtype of cancer but also to learn about the underlying characteristics of the cancer subtypes. Of interest is differentiating the structure of the gene networks in the cancer subtypes as a means of identifying  biologically significant differences that explain the variations between the subtypes. The modelling and inferential challenges are related to determining the number of components, as well as estimating the underlying graph for each component. We propose novel hierarchical extensions of our methods using finite mixture models and generalizations for infinite mixtures using Dirichlet process priors, which to the best of our knowledge, has not been addressed previously in the literature.

In Section 3 we describe the application of the proposed methods to  real data sets and comparisons with existing methods from the literature. In Section 4 we report results from simulations to assess the operating characteristics of our methods. In Section 5 we extend our method to develop a finite mixture of adaptive graphical models to handle clustered data. In Section 6 we incorporated the modelling framework into a Dirichlet process to automatically cluster the data when the number of cluster is unknown  and show the application of the model to a gene expression data set and appropriate simulations to validate the method. We provide a discussion and conclusion in Section 7, and an appendix.

\section{Bayesian graphical lasso selection model}
{To formalize the notation, let $\bm{Y}_{p\times n} = (\bm{Y}_1,\ldots ,\bm{Y}_n)$ be a $p \times n$ matrix with $n$ independent samples and $p$ variates, where each sample $\bm{Y}_i = (Y_i^{(1)},\ldots,Y_i^{(p)})$ is a $p$ dimensional vector corresponding to the $p$ variates; each sample $\bm{Y}_i$ comes from a multivariate normal distribution with mean $\bm \mu$; and the covariance matrix between the $p$ variates is $\Sigma$.  We have n samples with covariance matrix $\sigma^2\bm{I}_n$, which implies n independent samples with variance $\sigma^2$. This can be formalized as follows: $\bm{Y}$ follows a matrix normal distribution $\bm{N}(\bm\mu,\bm{\Sigma},\sigma^2\bm{I}_n)$ with mean $\bm\mu$ and nonsingular covariance matrix $\bm\Sigma$ between the $p$ variates $(Y^{(1)},\ldots,Y^{(p)})$ and $\sigma^2$ is the variance of the samples. This is identical to the multivariate normal distribution $\bm{MVN}(\bm\mu_v,\bm{\Sigma}\otimes\sigma^2\bm{I}_n)$ where $\bm\mu_v$ is the vectorized form of matrix $\bm\mu$ and $\otimes $ is the Kronecker product. }

 Given a random sample $\bm{Y}_1,\ldots ,\bm{Y}_n$ , we wish to estimate the precision  matrix $\bm{\Omega} = \bm{\Sigma}^{-1} $. The maximum likelihood estimator of $(\bm\mu,\bm\Sigma)$ is $(\bar{\bm{Y}},\bar{\bm{V}})$ where $\bar{\bm{V}}=\frac{1}{n}\sum_{i=1}^{n}{(\bm{Y}_{i}-\bar{\bm{Y}}){(\bm{Y}_{i}-\bar{\bm{Y}})}^{T}}$. The commonly used sample covariance matrix is $\hat{\bm{S}}=n\bar{\bm{V}}/(n-1)$. The precision matrix $\bm\Omega$ can be estimated by ${\bar{\bm{V}}}^{-1}$ or $\hat{\bm{S}}^{-1}$.

 However, if the dimension is $p$, we need to estimate $p(p+1)/2$ numbers of unknown parameters, which even for a moderate size $p$, might lead to unstable estimates of $\bm\Omega$. In addition, given that our main aim is to explore the conditional relationships among the variables, our main  interest is the identification of zero entries in the precision matrix because a zero entry $\Omega_{ij} = 0$ indicates conditional independence between the two covariates $Y^{(i)}$ and $Y^{(j)}$, given all other covariates.  We propose different  priors over $\bm\Omega$ to explore these zero entries. Here and throughout the paper we follow the notation that $\theta_1|\theta_2 $ represents the conditional distribution of the random variable $\theta_1$ given $\theta_2$, so that the likelihood of the Gaussian graphical model is  written as
\begin{eqnarray*}
\bm{Y}|G & \sim & \bm{N}(\bm{0},\bm{\Omega^{-1}},\sigma^2\bm{I}_n) \\
 & = & (2\pi \sigma^2)^{-\frac{np}{2}}|\bm{\Omega}|^{\frac{n}{2}}  exp\{-\frac{1}{2\sigma^2}tr\{\bm{\Omega Y Y^T}\} \}. 
\end{eqnarray*}

{Instead of modelling  the entire $p\times p$ precision matrix, $\bm\Omega$, we explore local dependencies by
 breaking the model down into components. In our modelling framework, we work directly with standard deviations and a correlation matrix following the {\it separation} strategy of \citet{barnard2000} that do not correspond to any particular type of parameterization (e.g., the Cholesky decomposition). Specification of a reasonable prior for the entire precision matrix is complicated because Wishart priors (or equivalents) are not fully general, but restrict the degrees of freedom of the partial standard deviations. Using this decomposition, prior beliefs on the partial standard deviations and correlations can be easily accommodated, as we show below.
}

To this end, we can parameterize the precision matrix as $\bm\Omega = \bm{S} \bm{C} \bm{S}$, where $\bm{S}$ is the diagonal matrix of standard deviations and $\bm{C}$ is the correlation matrix. The partial correlation coefficients $\rho_{ij}$ are related to $C_{ij}$ as 
\begin{eqnarray*}
\rho_{ij} = \frac{-\Omega_{ij}}{(\Omega_{ii}\Omega_{jj})^{\frac{1}{2}}} = -C_{ij}.
\end{eqnarray*}

In this setting, our primary construct of interest is $\bm{C}$, which we wish to model in an adaptive manner. One could model the elements of $\bm{C}$ directly using shrinkage priors such as Laplace priors, which gives us the Bayesian formulation of the graphical lasso models explored by \citet{mein06}, \citet{yuan2007} and \citet{friedman08}. However, using the Laplace priors leads to shrinkage of the partial correlation but does not set them exactly to zero, i.e., does not explicitly carry out selection, which is of essence in graphical models. To this end, we decompose the correlation matrix $\bm{C}$ as
 $$\bm{C} = \bm{A}\odot \bm{R},$$
where $\odot$ is the Hadamard operator that conducts the element-wise multiplication. This parameterization helps us to divide the problem into two parts: shrinkage and selection, using the respective elements of $\bm{R}$ and $\bm{A}$, with the following properties: (a) both $\bm{A}$ and $\bm{R}$ are symmetric matrices; (b)  the diagonal elements of both matrices are ones and, (c) the off-diagonal elements of the {\it selection matrix} $\bm{A}$ consist of binary random variables (0 or 1), whereas the off-diagonal elements of the {\it shrinkage matrix} $\bm{R}$ model the partial correlations between the variables with elements that lie between [-1, 1].  Moreover, $\bm{R}$ can thought of as a {\it pre-correlation matrix} which  may not be a positive definite matrix but we constrain the convoluted correlation matrix $\bm{C} = \bm{A}\odot \bm{R}$ to be positive definite. We do so by jointly modelling $\bm{A}$ and $\bm{R}$  as detailed below.

\subsection{ Joint prior specification of $\bm R$ and $\bm A$}

In order to achieve adaptive shrinkage of the partial correlations, we assign a Laplace prior to the off-diagonal elements of $\bm{R}$,  $R_{ij}$'s for $i<j$, where the 
Laplace prior is defined as 
$$f(R_{ij}|\tau_{ij})\propto \frac{1}{2\tau_{ij}}exp(-\frac{|R_{ij}|}{\tau_{ij}}),
$$ 
with each individual element having its own scale parameter, $\tau_{ij}$, that controls the level of sparsity. As discussed previously, Laplace priors have been widely used for shrinkage applications.

Since $\bm{A}$ is the selection matrix that performs the variable selection on the elements of the matrix $\bm{R}$, it thus consists of only binary variables with  the off-diagonal elements being either zeros or ones. The most general prior is an exchangeable Bernoulli prior on the off-diagonal elements of $\bm{A}$, given as
$$A_{ij}|q_{ij} \sim  \text{Bernoulli}(q_{ij}) ,i < j,$$
where $q_{ij}$ is the probability that the $ij^{th}$ element will be selected as 1.

To specify a joint prior for $\bm{A}$ and $\bm{R}$, we have to satisfy the constraint that $\bm{C} = \bm{A}\odot \bm{R}$ is positive definite. Hence, the joint prior can be expressed as$$R_{ij},A_{ij}|\tau_{ij},q_{ij}\sim \text{Laplace}(0,\tau_{ij})  \text{Bernoulli}(q_{ij})I(\bm{C}\in \mathbb{C}_p),$$ where $-1\leq R_{ij}\leq 1$, $0\leq q_{ij}\leq 1$ and $I(\bm{C}\in \mathbb{C}_p)=1$ if $\bm{C}$ is a correlation matrix and is 0 otherwise. Therefore, we ensure the positive definiteness of $\bm C$ using plausible values of $\bm{A}$ and $\bm{R}$. That way,  $R_{ij}$ and $A_{ij}$ are dependent through the indicator function. The joint prior of $\bm{A}$ and $\bm{R}$ can be specified as $$\bm{A},\bm{R}|\bm{\tau},\bm{q}\sim \prod_{i<j}\text{Laplace}(0,\tau_{ij}) \text{Bernoulli}(q_{ij}) I(\bm{C}\in \mathbb{C}_p) $$ where $\bm{\tau}$ and $\bm{q}$ are the vectors containing $\tau_{ij}$ and $q_{ij}$ values respectively. The full specification of the constraints on the $C_{ij}$'s to ensure the positive definiteness are discussed in Appendix 1.

\begin{figure}[h!]
\subfigure[Prior for p=3]{\includegraphics[scale =.5,angle=0]{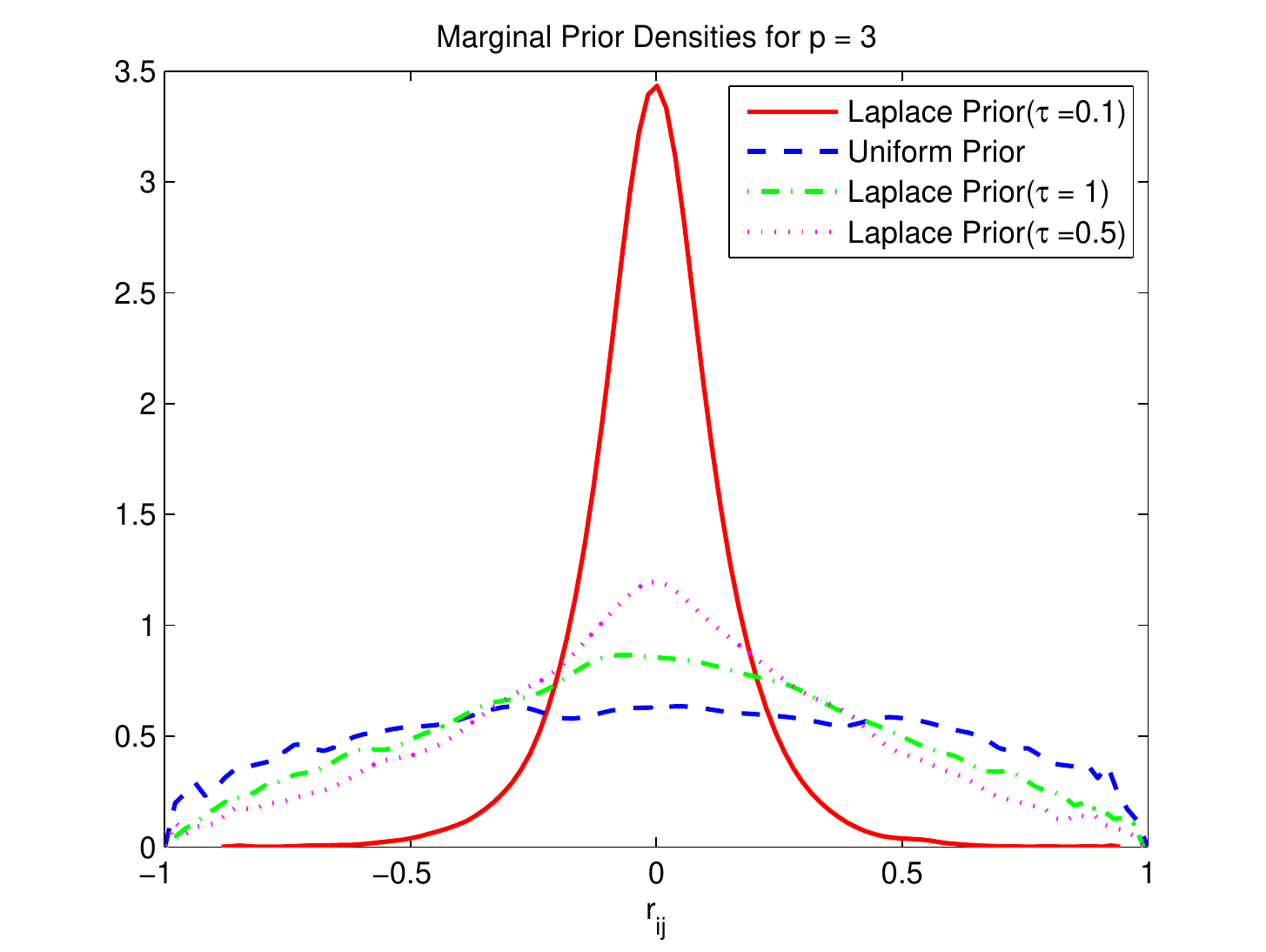}}
\subfigure[Prior for p=10]{\includegraphics[scale =.5,angle=0]{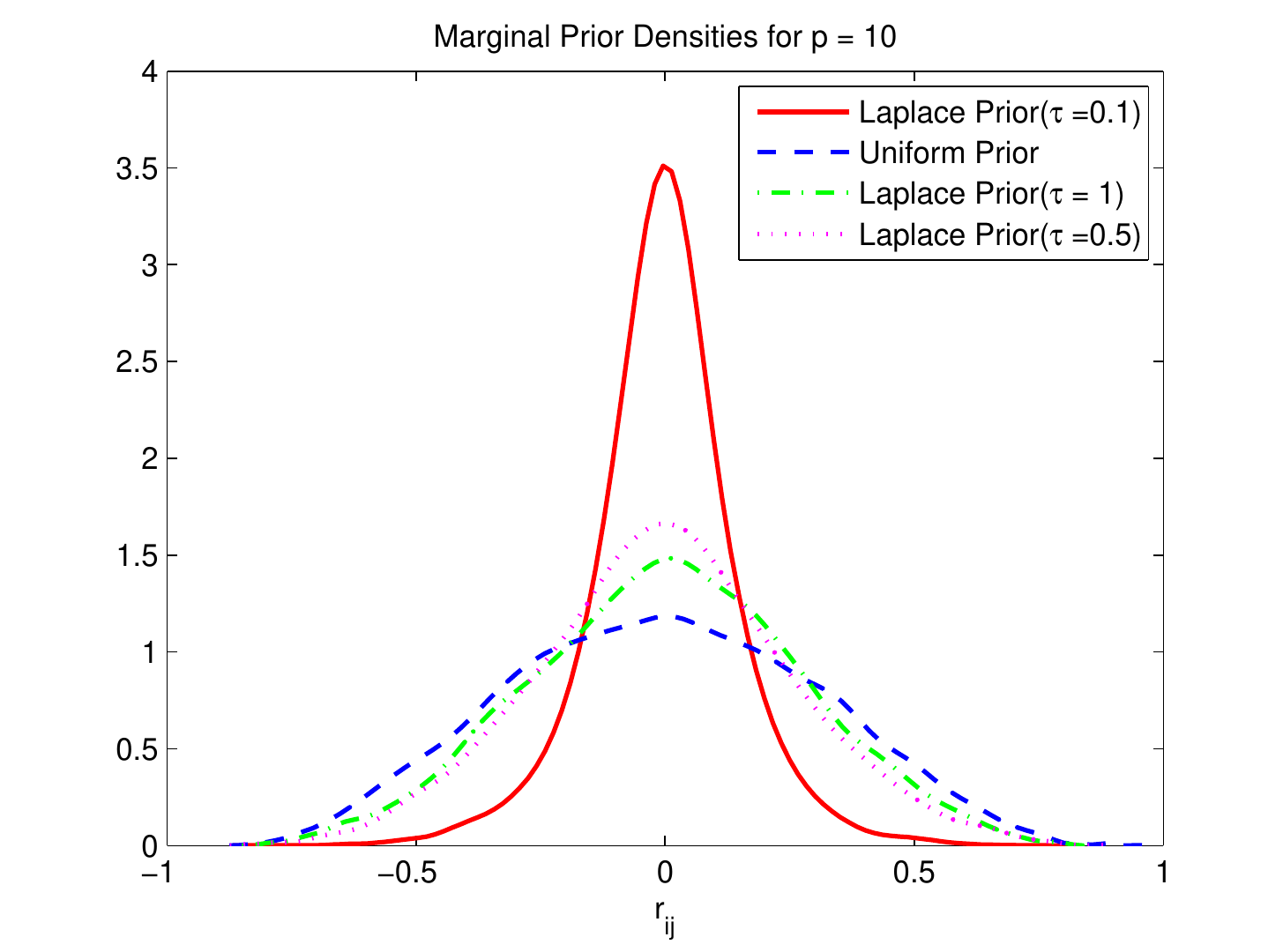}}
\subfigure[Prior for p=20]{ \includegraphics[scale =.5,angle=0]{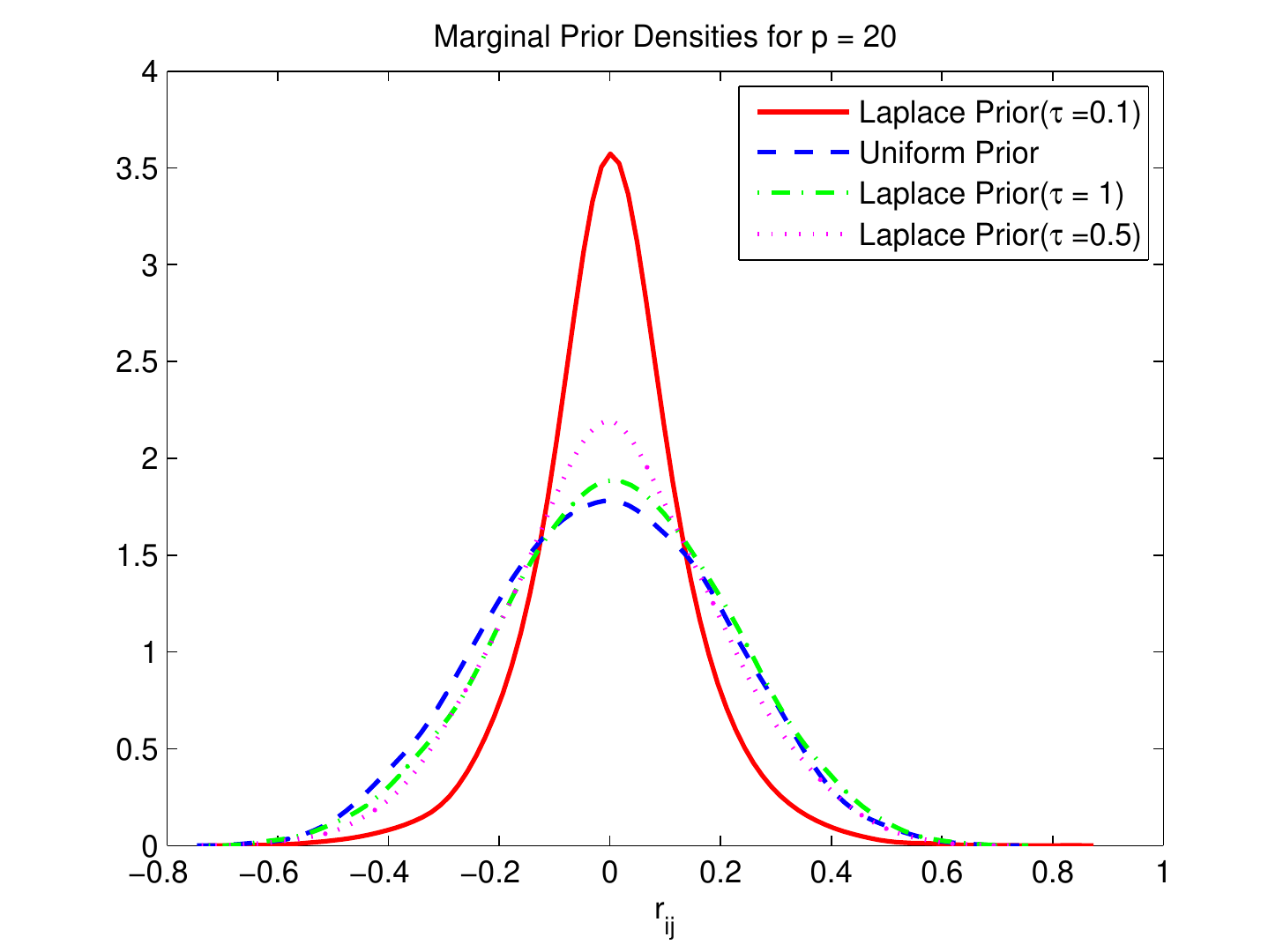}}
\caption{ Shown here are the marginal priors on $R_{ij}$ for different values of $\tau=1,.5,.1$ and compared with the uniform prior.}
\label{simuprior}
\end{figure}

 $R_{ij}$s are not independent under this prior specification due to the constraint of positive definiteness of $\bm{C}$ through the indicator function. To observe the marginal prior distribution of $R_{ij}$, we simulate from this joint prior distribution by fixing $A_{ij}=1$ for all $i,j$ so that $\bm C$ is identical to $\bm R$. Figure~\ref{simuprior} shows marginal distributions of some of the $r_{ij}$'s with different values of $\tau_{ij}$'s when $p=$3, 10 and 20.  In addition, we have also plotted the marginal prior distribution arising from a constrained uniform distribution  as suggested by \citet{barnard2000}. It is clear from the figures that the marginal prior on individual correlations under the joint Laplace prior shrinks more tightly towards zero compared to the joint uniform prior. Furthermore, the effect of dimension $p$ is negligible on Laplace prior distribution for small $\tau_{ij}$ values.

Thus in this setting, the shrinkage parameter $\tau_{ij}$  controls the degree of sparsity, i.e., determines the degree to which the  $ij^{th}$ element of $\bm{R}$  will be shrunk towards zero. Accordingly, we treat this $\tau_{ij}$ as an unknown parameter and estimate them adaptively using the data. We assign an exchangeable  inverse gamma prior as
$$\tau_{ij} \sim  IG(e,f) ,i <j,$$ where $(e,f)$ are the shape and scale parameters, respectively. Note that if we set $\tau_{ij} = \tau ~~ \forall i,j $ along with $\bm{A} = \bm{1}_n$ (i.e., a matrix of all 1's), this gives rise to the special case of the Bayesian version of the graphical lasso of \citet{friedman08} and  \citet{yuan2007},  where the single penalty parameter ($\tau$) controls the sparsity of the graph and is estimated via cross-validation or by using a criterion similar to the Bayesian information criterion (BIC). By allowing the penalty parameter to vary locally for each node, we allow for additional flexibility, which has been shown to result in better properties than those of the lasso prior and which also satisfies the oracle property (consistent model selection), as shown by \citet{griffin} in the variable selection context. This fact is also illustrated in our data analysis (Section 3) and simulations studies (Section 4).

Furthermore, $q_{ij}$  is assigned a beta prior as
$$q_{ij} \sim  \text{Beta}(a,b) ,i < j.$$ In this construction the hyper-parameters $q_{ij}$ control the probability that the $ij^{th}$ element will be selected as a non-zero element. To evaluate a  highly sparse model the hyper-parameters should be specified such that the beta distribution is skewed towards zero, and for a dense model the hyper-parameters should be specified such that the beta distribution is skewed towards one. Furthermore, prior beliefs about the existence of edges can be incorporated at this stage of the hierarchy by giving greater weights to important edges while down-weighting redundant edges.

 In conclusion, the joint specification of $\bm{A}$ and $\bm{R}$ above gives us the {\it graphical lasso selection model} that performs simultaneous  shrinkage and selection.  To complete the hierarchical specification of the graphical lasso selection, 
we use an inverse gamma prior on the inverse of the partial standard deviations $S_i$:
$$S_i\sim IG(g,h) , i = 1,2,\ldots,p.$$

The complete hierarchical model can be succinctly summarized as
\begin{eqnarray*}
\bm{Y}|\bm{\Omega},\sigma^2&\sim &\bm{N}(\bm{0},\bm{\Omega}^{-1},\sigma^2\bm{I}_n)\\
\bm{\Omega}&=&\bm{S}(\bm{A}\odot \bm{R})\bm{S} \\
\bm{A},\bm{R}|\bm{\tau},\bm{Q}&\sim& \prod_{i<j}\text{Laplace}(0,\tau_{ij}) \text{Bernoulli}(q_{ij}) I(\bm{C}\in \mathbb{C}_p) \\
\tau_{ij} &\sim & IG(e,f) ,i < j \\
q_{ij} &\sim& \text{Beta}(a,b) ,i<j\\
S_i&\sim &IG(g,h)\\
\sigma^2&\sim& IG(k,l),
\end{eqnarray*}
where $i=1,\ldots,p$, $j=1,\ldots,p$  and  $\odot$ is the Hadamard product.  The posterior conditionals and sampling methods are detailed in  section 1 of supplementary material.

\section{ Data example} \label{sec:enron}

We take a motivating example from a stock market dataset \citet{Liechty03}, which has been used by the finance community to group and analyze companies according to their areas of operation.  This grouping requires knowledge of the companies and is determined by people who are experts in the field. Grouping companies according to the services or products they offer may be complicated by companies redirecting their efforts, e.g., in response to changing economic situations or consumer demands.  


Enron was a company that provided a good illustration of this type of change. Enron began as an energy company, but changed its business focus and transformed itself into a finance company. It was not known whether Enron provided more service to energy clients or to finance clients; therefore, the category into which Enron fit was uncertain. One approach to resolving this uncertainty is to examine the behavior of a company’s stock to determine its primary service.  We undertook such an analysis using  the same data set that was used by \citet{Liechty03}, which consists of data on nine companies. Four of the companies were known to provide energy services, four were known to provide financial services, and the ninth was Enron. The energy companies were Reliant, Chevron, British Petroleum and Exxon. The finance companies were Citi-Bank, Lehman Brothers, Merrill Lynch and Bank of America. The data included monthly stock data for each company over a period of 73 months. This example is also motivated by the need for accurate estimates of pairwise correlations of assets in dynamic portfolio-selection problems. Graphical models offer a potent tool for regularization and stabilization of these estimates, leading to portfolios with the potential to uniformly dominate their traditional counterparts in terms of risk, transaction costs, and overall profitability.

We report the best graphs supported by the data by computing the  posterior probabilities for  the graphs using the following scheme. The Markov Chain Monte Carlo samples obtained from the analysis explore the distribution of possible graphical configurations suggested by the data, with each configuration represented by the selection matrix $\bm{A}$ encoding the indicators of the possible edges.  To explore the space of valid graphs, we follow the strategy of selecting the model with the highest marginal posterior probability over the space of all possible graphs. We obtain the Monte-Carlo estimates of these posterior probabilities by counting the proportion of Markov Chain Monte Carlo samples to have the specific graphical structure. Hence, if $I(\bm{A}=\bm{A}^*)$ denote the indicator function for the graphical model $\bm{A}=\bm{A}^*$ , then the ergodic average or the Monte Carlo frequency estimator of this model $\bm{A}^*$ is given by
\begin{equation*} 
\pi(\bm{A}^*|\bm{Y}) =\frac{1}{K} \sum_{b=1}^{K} I(\bm{A}_b=\bm{A}^*),
\end{equation*} 
where $\bm{A_b}$ is graphical model visited on the $b^{th}$ Markov Chain Monte Carlo draw and $K$ is the total number of  draws from the Markov chain.

\begin{figure}[h!]
\subfigure[$P(model|data) = 0.0736$]{\includegraphics[scale =.85,angle=0]{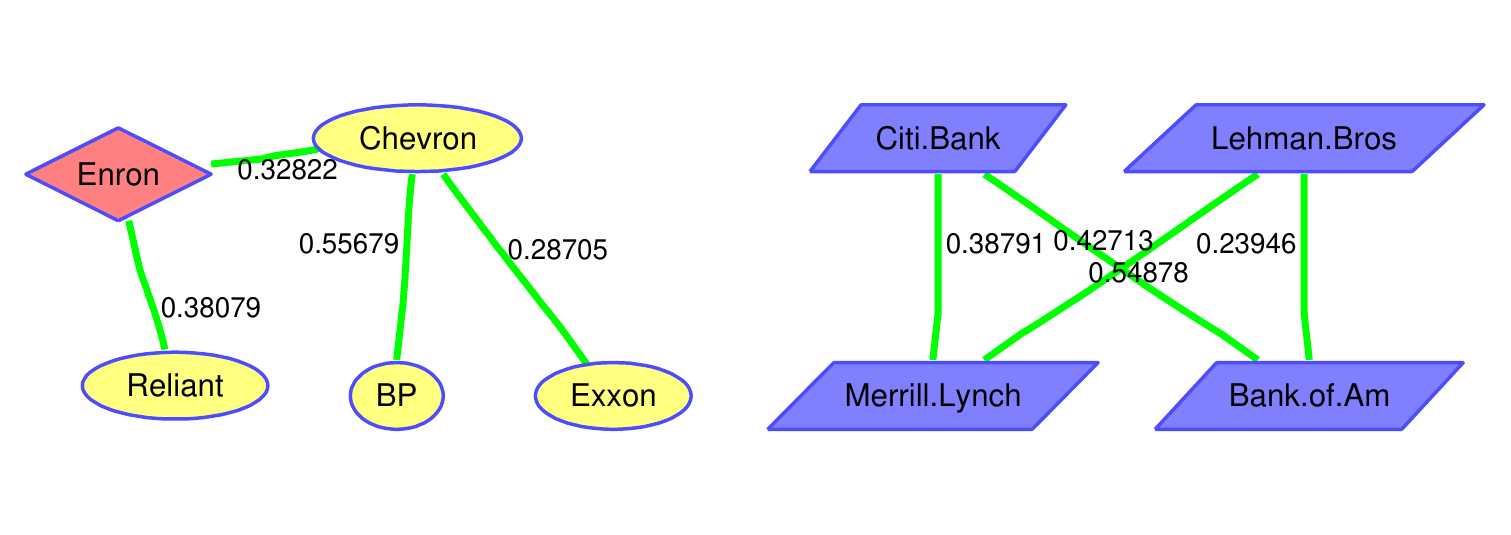}\label{bestenr}}\\
\subfigure[$P(model|data) = 0.0602$]{\includegraphics[scale =.85,angle=0]{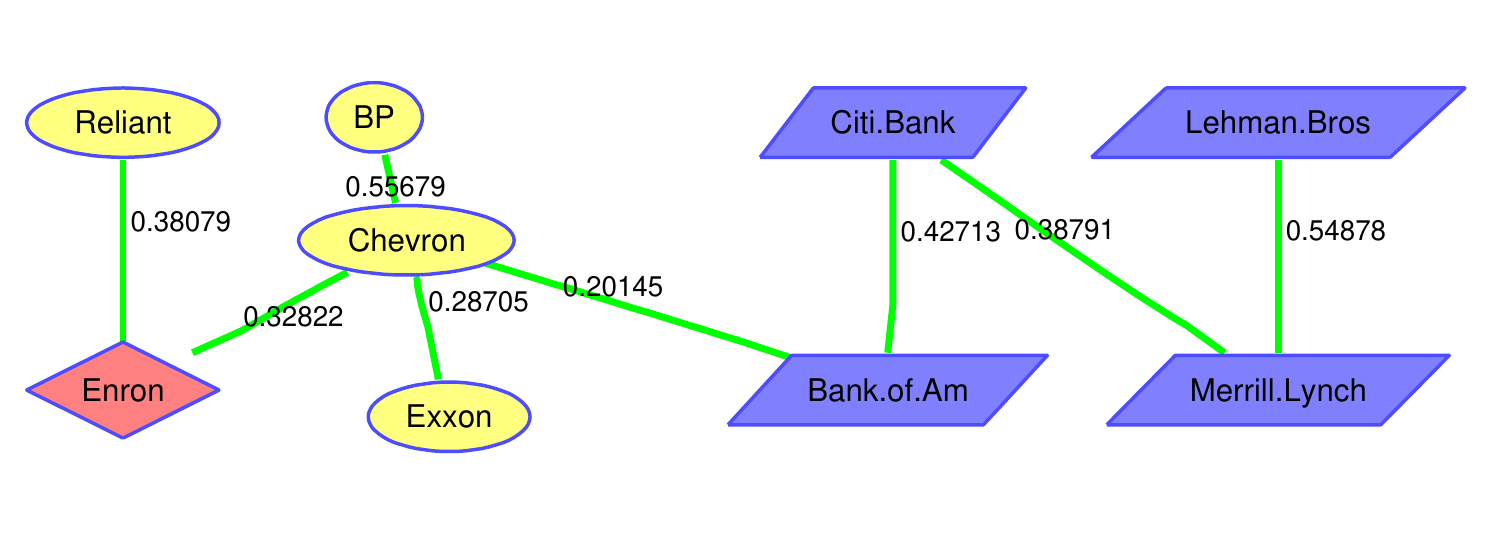}}\\
\subfigure[$P(model|data) =0.0590$]{ \includegraphics[scale =.85,angle=0]{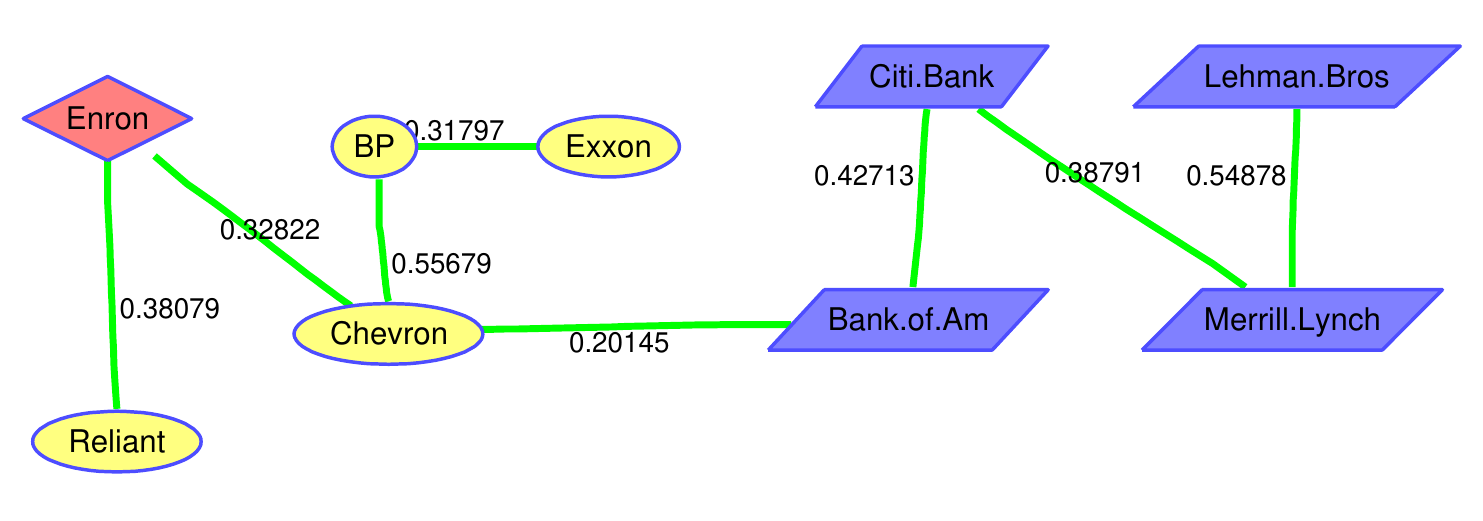}}
\caption{ Shown here are the top three graphical models for the stock market data, sorted by the marginal posterior probabilities of the models.}
\label{enron}
\end{figure}

{ The top three graphs identified using our lasso selection model  are   shown in Figure~\ref{enron} sorted by the posterior probabilities. It is clear from the illustrated network (e.g  Figure ~\ref{bestenr}) that Enron is grouped with the energy companies and was not successful, in terms of stock performance, in transitioning from an energy company to a finance company. \citet{Liechty03} also found  Enron to be more closely related to the energy companies than the finance companies.
}

\begin{figure}[h!]
\subfigure[Glasso]{\includegraphics[scale =.55,angle=0]{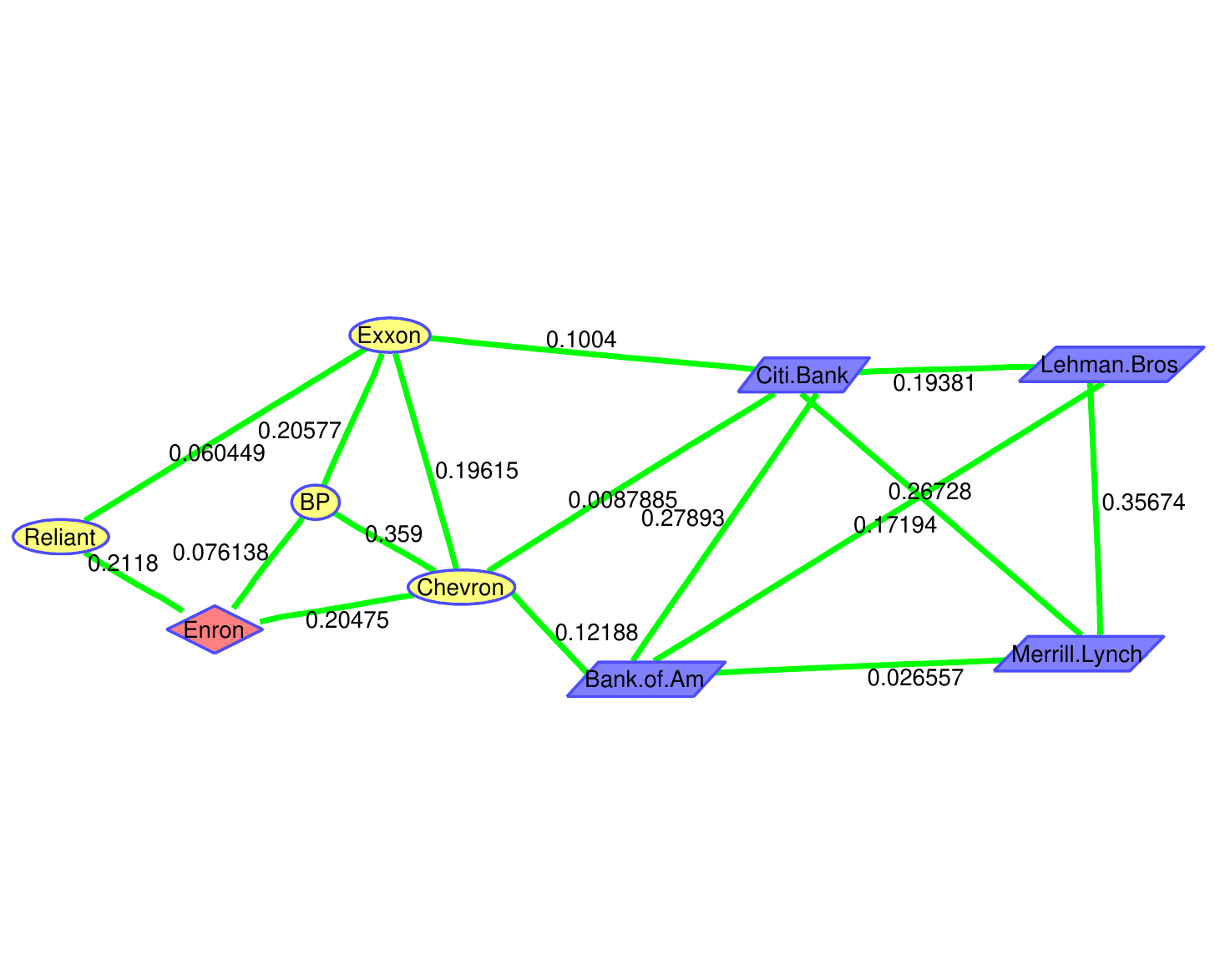}}
\subfigure[MB]{\includegraphics[scale =.55,angle=0]{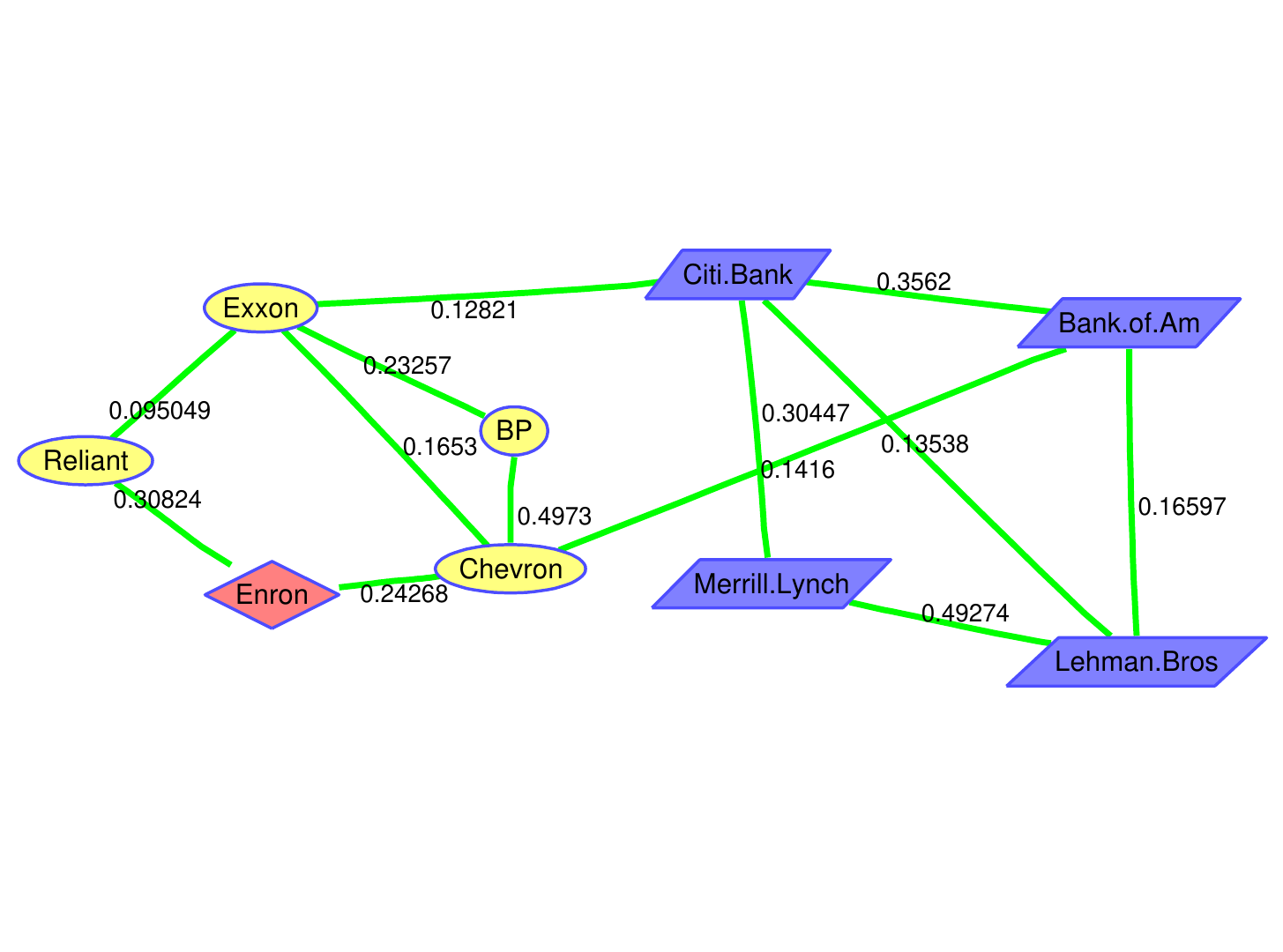}}
\caption{ The graphical models for the stock market data obtained using (a) the glasso method and (b) the MB method. }
\label{enrontest}
\end{figure}

For comparison with our proposed method, we selected two methods that use L1-regularization and are similar to our approach using Laplace priors: the ``glasso" approach of \citet{friedman08} and the method (``MB") proposed by \citet{mein06}. As both approaches are frequentist, hence they incorporate no notion of marginal likelihoods and posterior probabilities, we used 
prediction performance to compare the methods. We split the 73-month data sample into a 60-month training set
and a 13-month prediction set. Using the training set to find the top 10 graphs (where top graphs are ranked by marginal posterior probabilities), we applied the Bayesian lasso selection model and found the estimates of the precision matrix for each  graph.  We then predicted the stock value of each sample of the test set given all other stocks for each of the test samples and averaged them over the 10 graphs -- thus employing Bayesian model mixing. For the glasso and MB methods, we used the estimate for the precision matrix derived by these methods to predict the test samples  using $\rho = 0.1$, where $\rho$ is the tuning parameter for the lasso penalty in both methods. For the sake of a fair comparison of the frequentist methods, we also included a Bayesian model with a single penalty parameter, making  $\tau_{ij} = \tau$ and $q_{ij} = q$ to make it equivalent to the frequentist models with a single penalty parameter.  
The results are shown in Table~\ref{pred}. 
\begin{table}
  \caption{\label{pred}Predictive squared error comparison for Enron stock data}
    \fbox{
    \begin{tabular}{*{4}{c}}
            Bayesian lasso selection& Bayesian lasso (single penalty)  & glasso & MB \\
            \hline
   30.6764 &31.9765& 32.1968 & 32.7445 
    \end{tabular}}
\end{table}
We can see that the Bayesian lasso selection model has the lowest (better) predictive squared error compared to the frequentist methods, thus showing how Bayesian model mixing can help improve prediction accuracy. Of interest is that the performance of the Bayesian lasso model with the single penalty parameter was worse than that of the lasso selection model with a locally varying penalty, and its prediction performance was close to those of the glasso and MB methods.  We show the graphs derived from the glasso and MB methods in Figure~\ref{enrontest}. The inferences are similar using these approaches in the sense that Enron is linked more with oil companies than  finance companies. However, these approaches show more connections than are shown in our selection models. Thus the methods seem to differ in imparting sparse solutions, with the Bayesian lasso selection models giving sparser outputs, which is reflected in the prediction performance.

In addition, we compared our graphical method to a simple cluster analysis to see how the companies cluster together in terms of their stock performance. We clustered the data using the model-based clustering software MCLUST (\citealp{mclust02}). We used the  ``VVV" parameterization to estimate the unconstrained covariance matrix for the data  and used BIC to find the optimal number of clusters. The optimal number of clusters found by BIC was one cluster, which grouped all nine companies together. In contrast, the graph with the highest posterior probability as determined by our method, Figure ~\ref{bestenr}, detected two distinct subgraphs, those of energy companies and finance companies, with Enron being connected to the energy companies. This clustering also appeared in the other graphs in Figure~\ref{enron}. In essence, cluster analysis missed this relationship and was unable to distinctly answer the scientific question that was posed.

\section{Simulations} \label{sec:sims}
In this section we use simulations to compare different methods of assessing the performance of the adaptive Bayesian graphical models. We include a comparison of the performances of a shrinkage model and a selection model, using a shrinkage model based on a structure similar to that of the selection model.  We simulate the following five types of precision matrices, which are listed in order of increasing structural complexity:
\begin{enumerate}
\item[1.] Identity matrix.
\item[2.] Banded diagonal matrix.
\item[3.] Block diagonal matrix.
\item[4.] Sparse unstructured matrix.
\item[5.] Dense unstructured matrix.
\end{enumerate}

\begin{figure}[h!]
\center
\subfigure[Identity Matrix]{\includegraphics[scale =.35,angle=0]{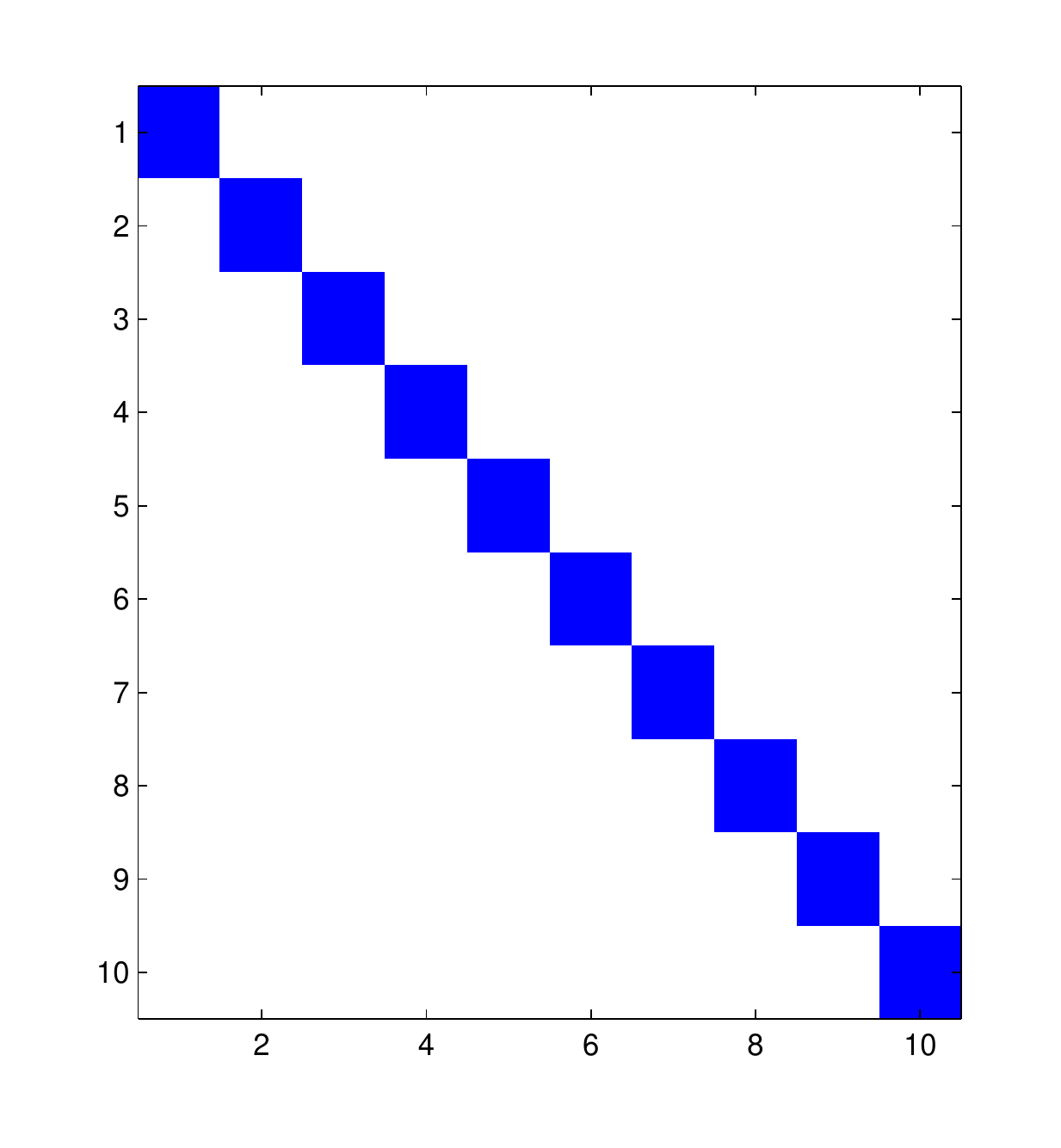}}
\subfigure[Banded Diagonal Matrix]{\includegraphics[scale =.35,angle=0]{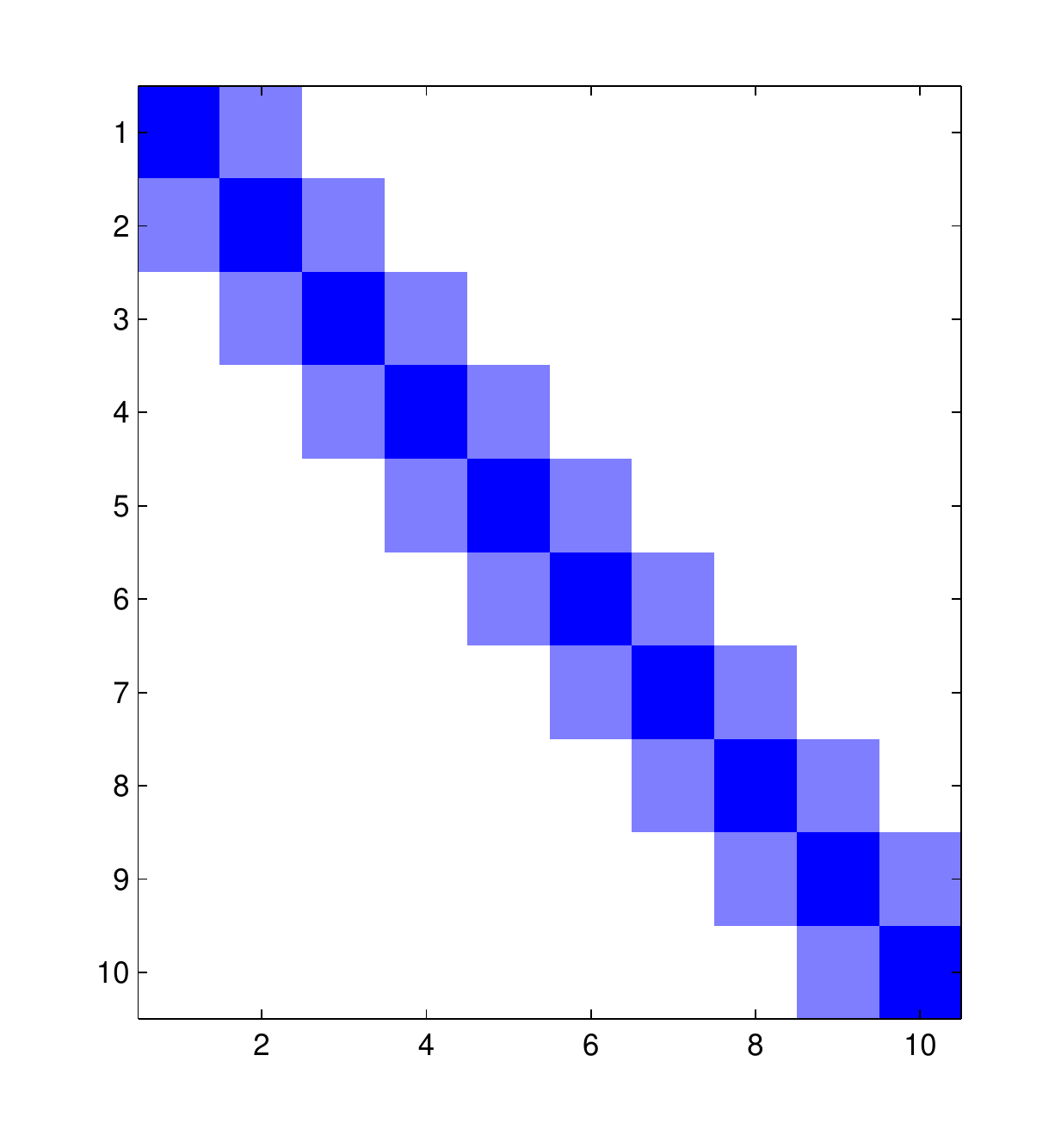}}
\subfigure[Block Diagonal Matrix]{\includegraphics[scale =.35,angle=0]{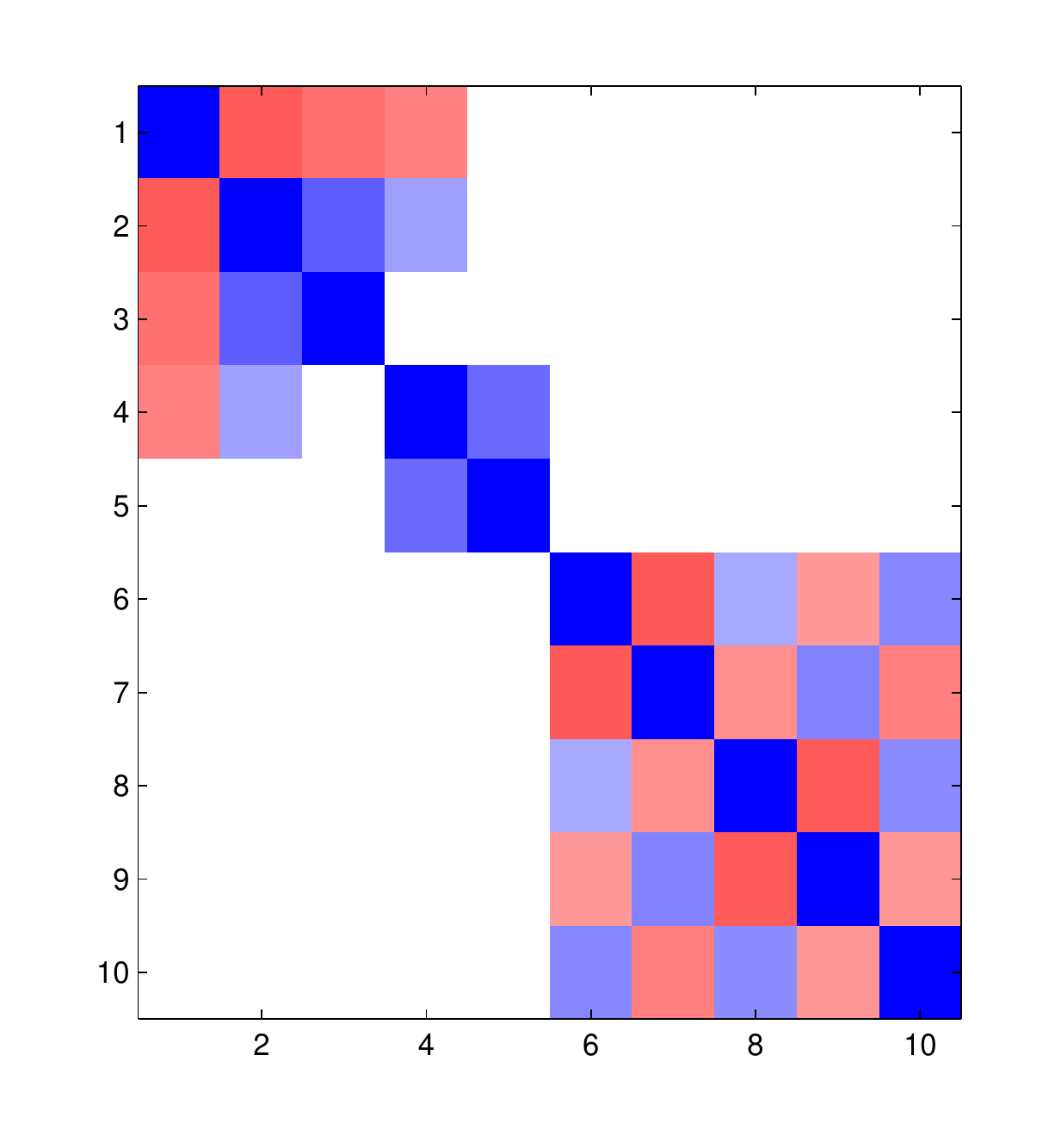}\label{blockmat}}\\
\subfigure[Sparse  Matrix]{\includegraphics[scale =.35,angle=0]{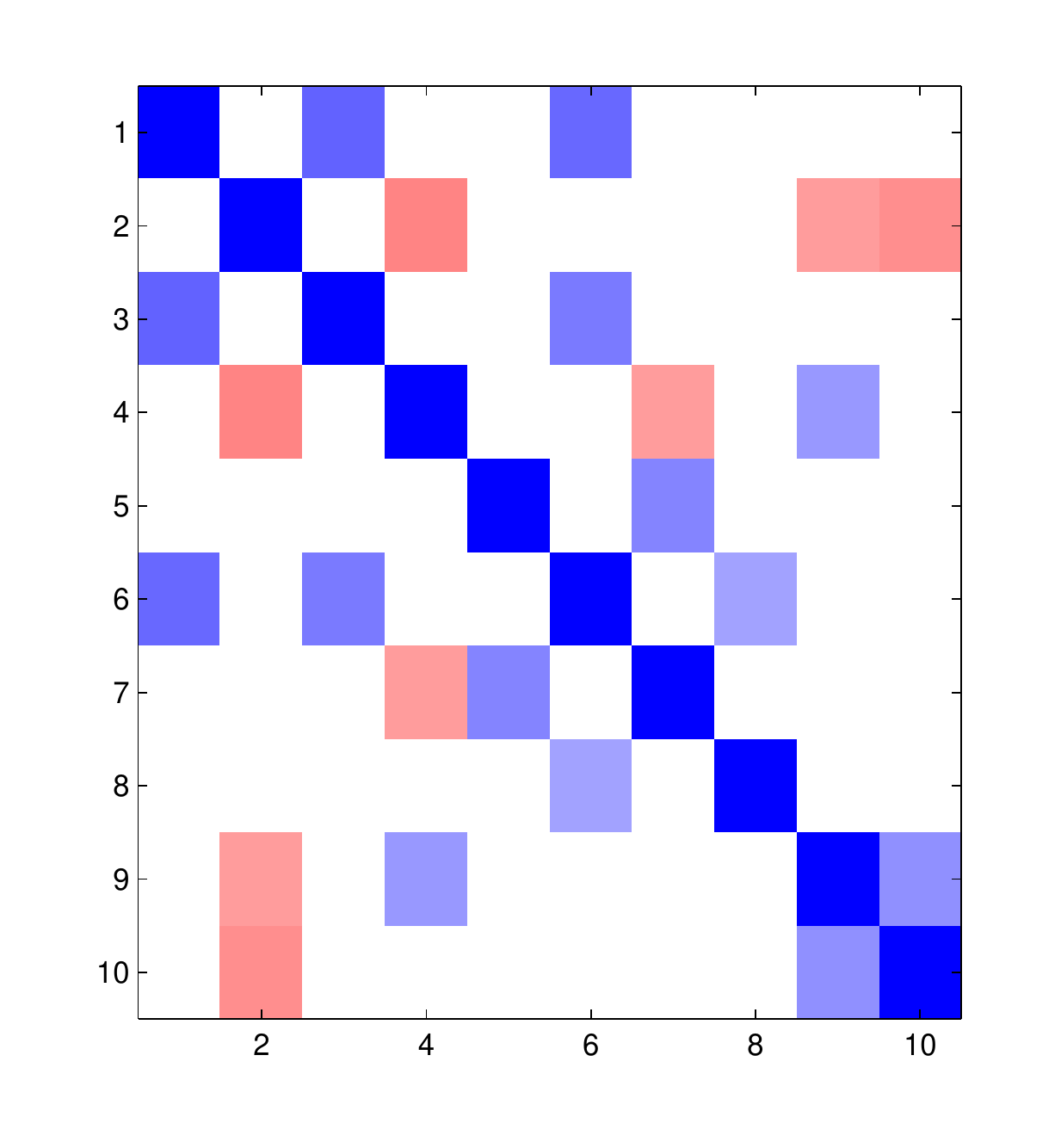}}
\subfigure[Dense  Matrix]{\includegraphics[scale =.35,angle=0]{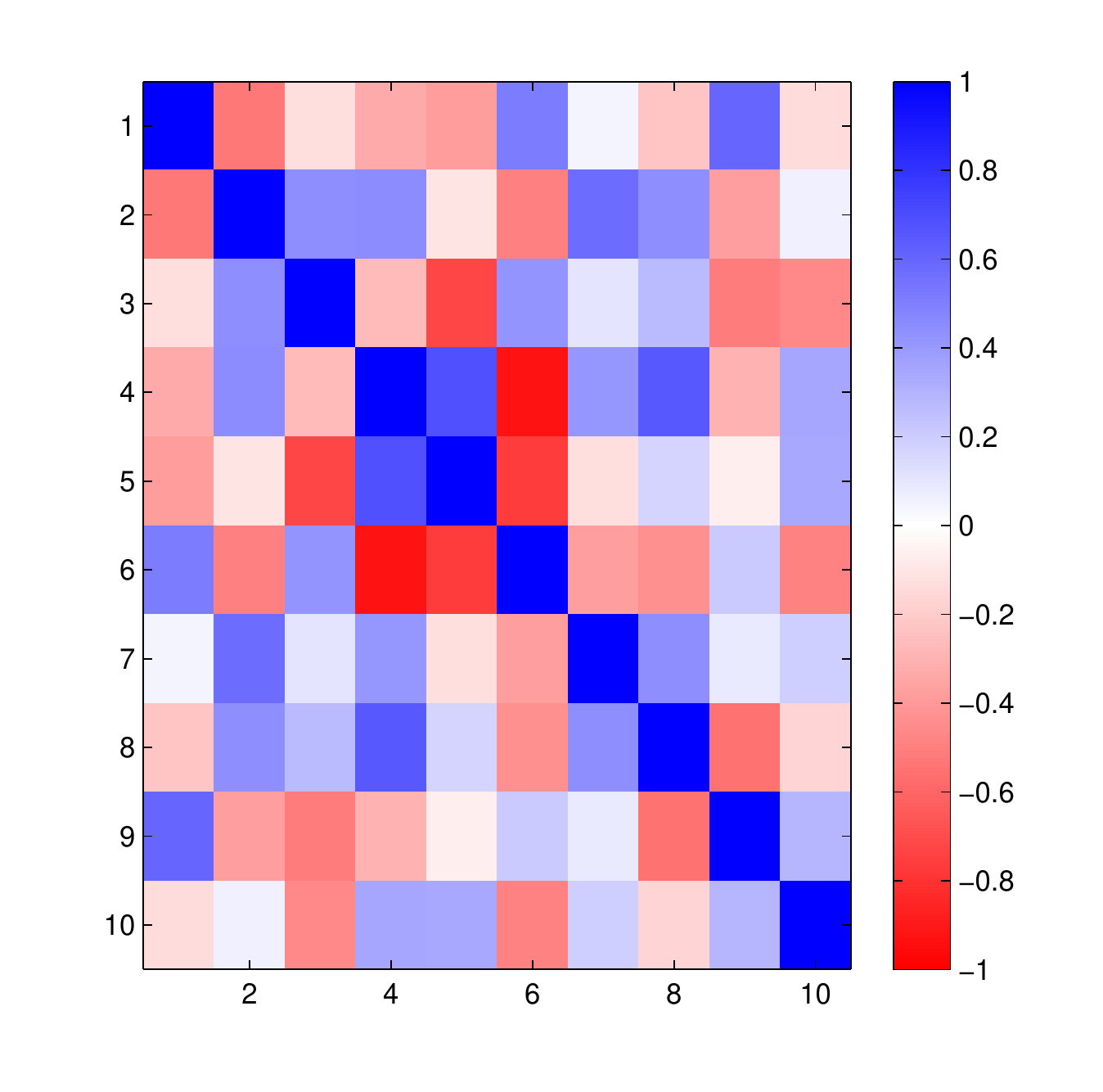}}
\caption{Simulated matrices for different types of structures of the precision matrix, based on the same colorbar for all the matrices. White indicates a zero in the precision matrix;  colored cells indicate non-zero elements.}
\label{simumat}
\end{figure}

An  identity matrix is a simple matrix with ones in its diagonal and zeros in its off-diagonal. A banded diagonal matrix is a tridiagonal matrix with ones in its diagonal and all the elements in the diagonals adjacent to the main diagonal set to 0.5. Before explaining simulations of more the complex matrix structures, we describe the process used for generating a random positive-definite correlation matrix. A random lower triangular matrix $\bm{L}$ was generated with ones in its diagonal and normal random numbers in its lower triangle. $ \bm{L}\bm{L}^T$ gave us a positive-definite matrix. The matrix was then factored as $\bm{Q}\bm{\Omega}\bm{Q}$, where $\bm{Q}$ is a diagonal matrix and $\bm\Omega$ is a correlation matrix with ones in its diagonal, which is the desired positive-definite correlation matrix.  A block diagonal matrix was generated as follows. Two positive-definite matrix correlation matrices of sizes $p-k$ and $k$ were generated, where k is a random number between $1$ and $p$. The correlation matrices were concatenated in the diagonals to create a matrix of size $p\times p$, as shown in figure~\ref{blockmat}. The sparse unstructured matrix was simulated as follows: Let $\bm{M} = \bm{V} + \delta \bm{I}_p$ where each off-diagonal entry in $\bm{V}$ is generated independently and equals a random number between $[-1,-.5]$ and $[.5,1]$ with probability $\pi$ or 0 and with probability $1- \pi$, all diagonal entries of $\bm{V}$ are zero and $\delta$ is chosen such that the resulting matrix is positive  definite. In the end we complete the factorization of $\bm{M}$ as $\bm{Q}\bm{\Omega}\bm{Q}$,  where $\bm{Q}$ is a diagonal matrix and $\bm\Omega$ is a correlation matrix with ones in its diagonal, which is the desired sparse positive-definite correlation matrix.  We can vary the sparsity of the matrices generated by changing the value of $\pi$.  We chose  $\pi = 0.1$ for the sparse unstructured matrix. The dense unstructured matrix is the full matrix that is a random positive-definite correlation matrix of size $p$ generated using the method described above. The simulated matrices for sizes $p=10$ and $n=25$ are shown in Figure~\ref{simumat}. The white blocks in the figure are the zeros in the matrices;  the colored blocks correspond to the magnitude of non-zero off-diagonal  elements in the matrices, as represented by the colorbar adjacent to Figure 3(e), in which red indicates a negative association and blue indicates a positive association.

As in Section \ref{sec:enron}, we compared our method with the glasso and MB methods because both methods use the L1-regularization and are closest to our approach using Laplace 
priors. We assessed the performance of these methods in terms of the Kullback-Leibler (K-L) loss. Both methods were implemented using the glasso package in R programming language. We implemented them using the Matlab-R link to call the functions in Matlab. It should be noted that both these methods are frequentist methods and give a point estimate for the precision matrix, whereas the Bayesian methods  also provide uncertainty estimates for the covariance matrix.  As a result, we compared the performance of the methods with respect to the final estimate of the precision matrix. 

\begin{figure}[h!]
\vspace{-1cm}
\subfigure[Identity Matrix]{\includegraphics[scale =.5,angle=0]{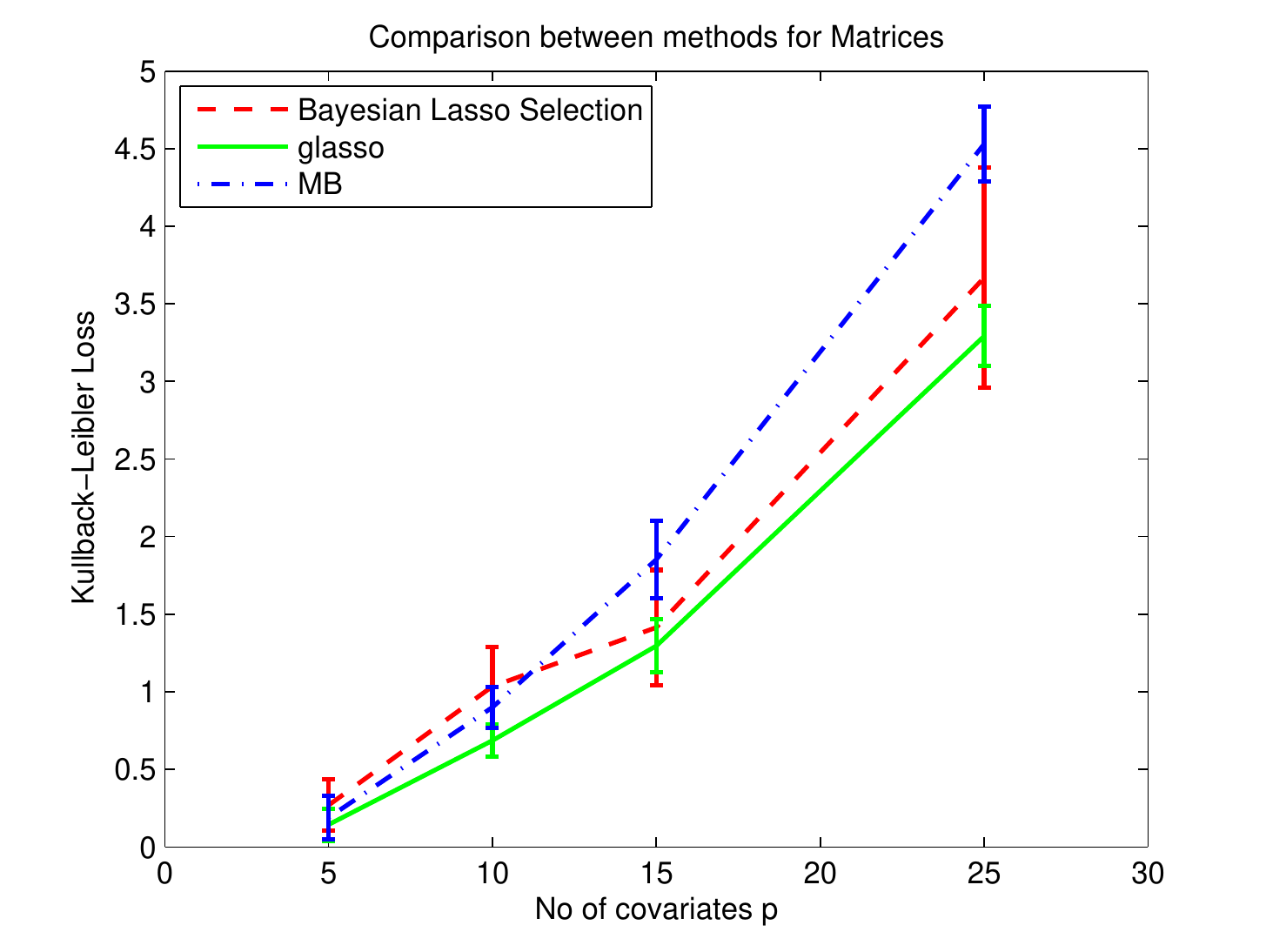}}
\subfigure[Banded Diagonal Matrix]{\includegraphics[scale =.5,angle=0]{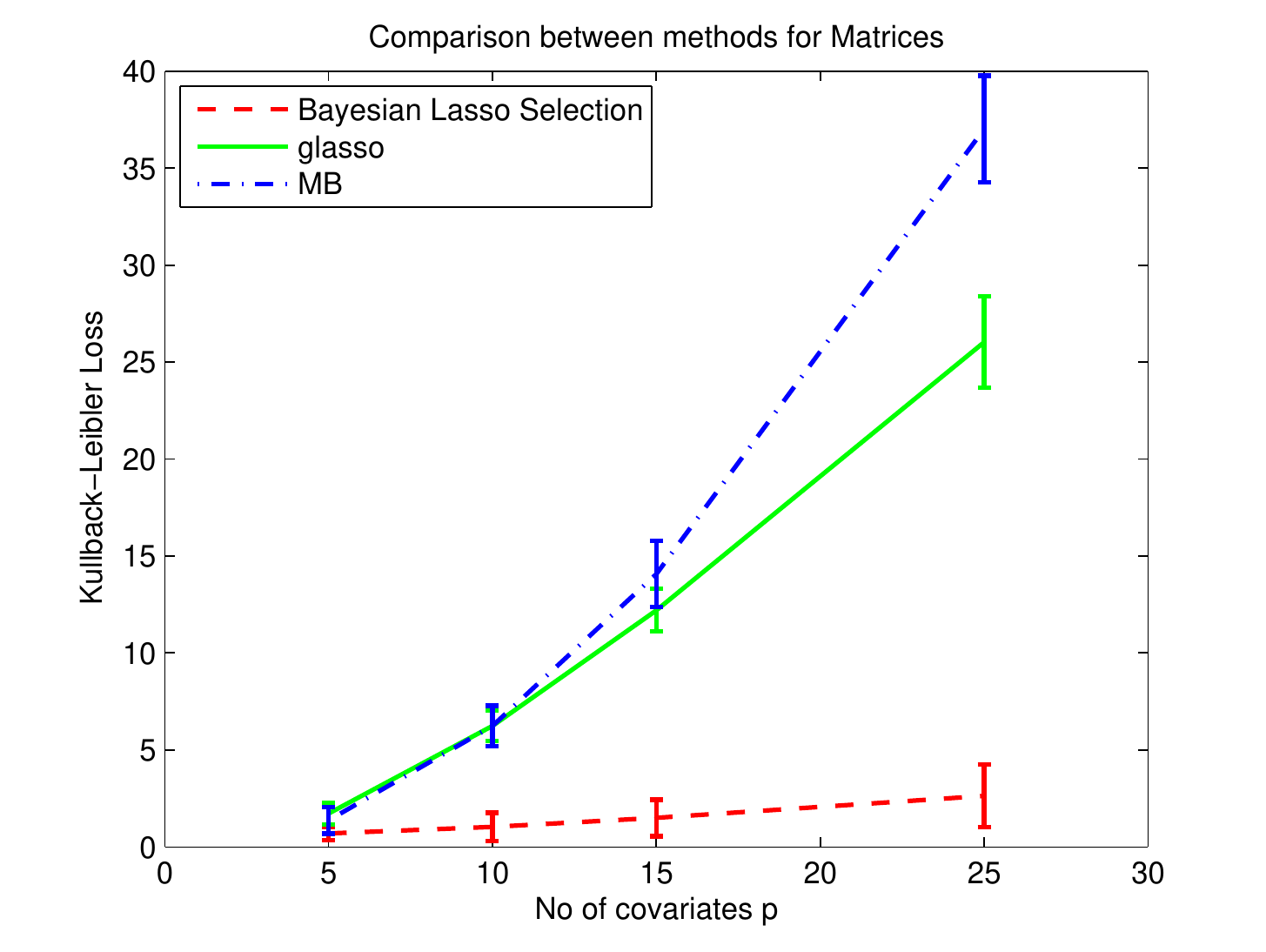}}
\subfigure[Block Diagonal Matrix]{\includegraphics[scale =.5,angle=0]{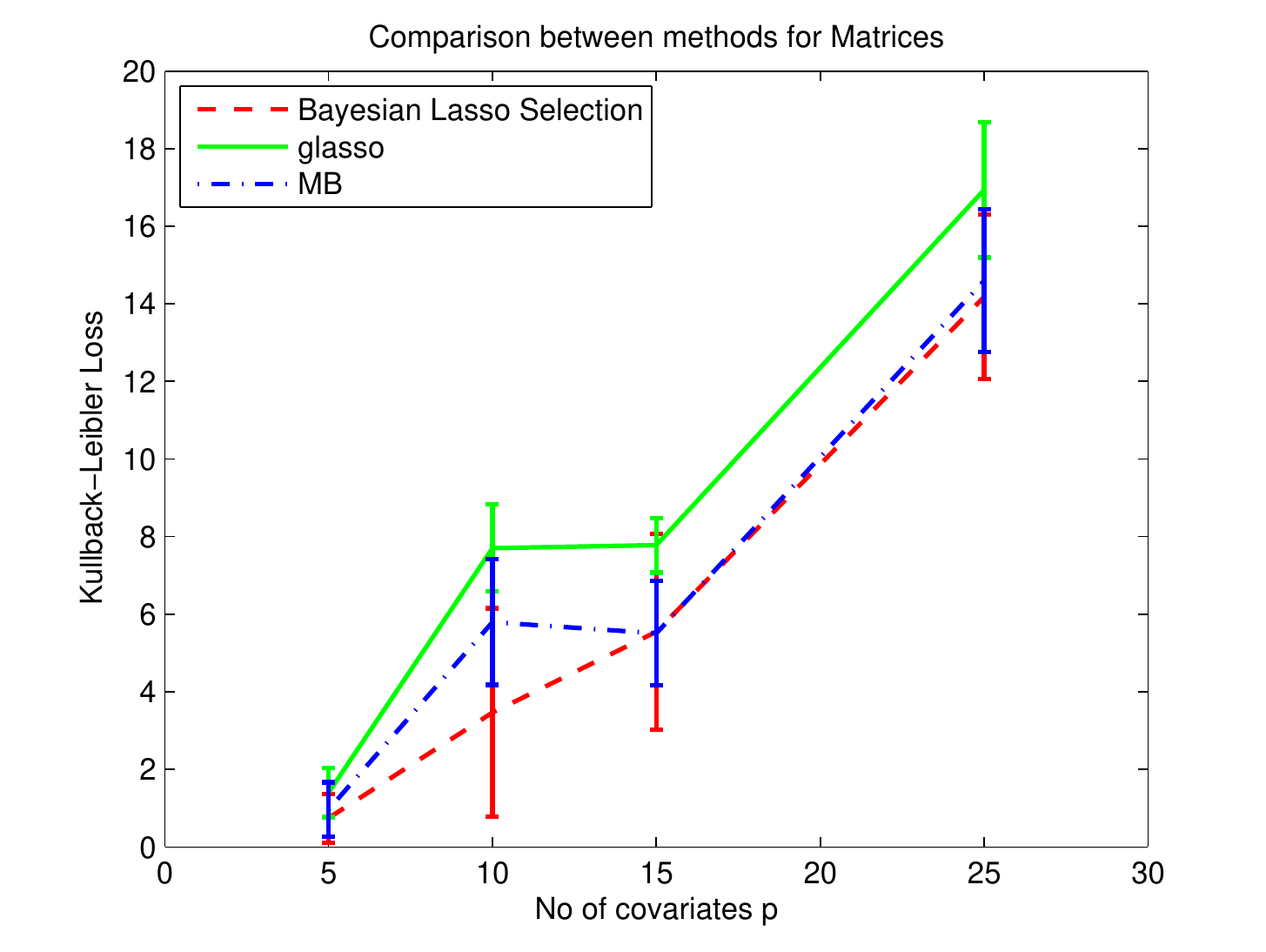}}
\subfigure[Sparse Matrix]{\includegraphics[scale =.5,angle=0]{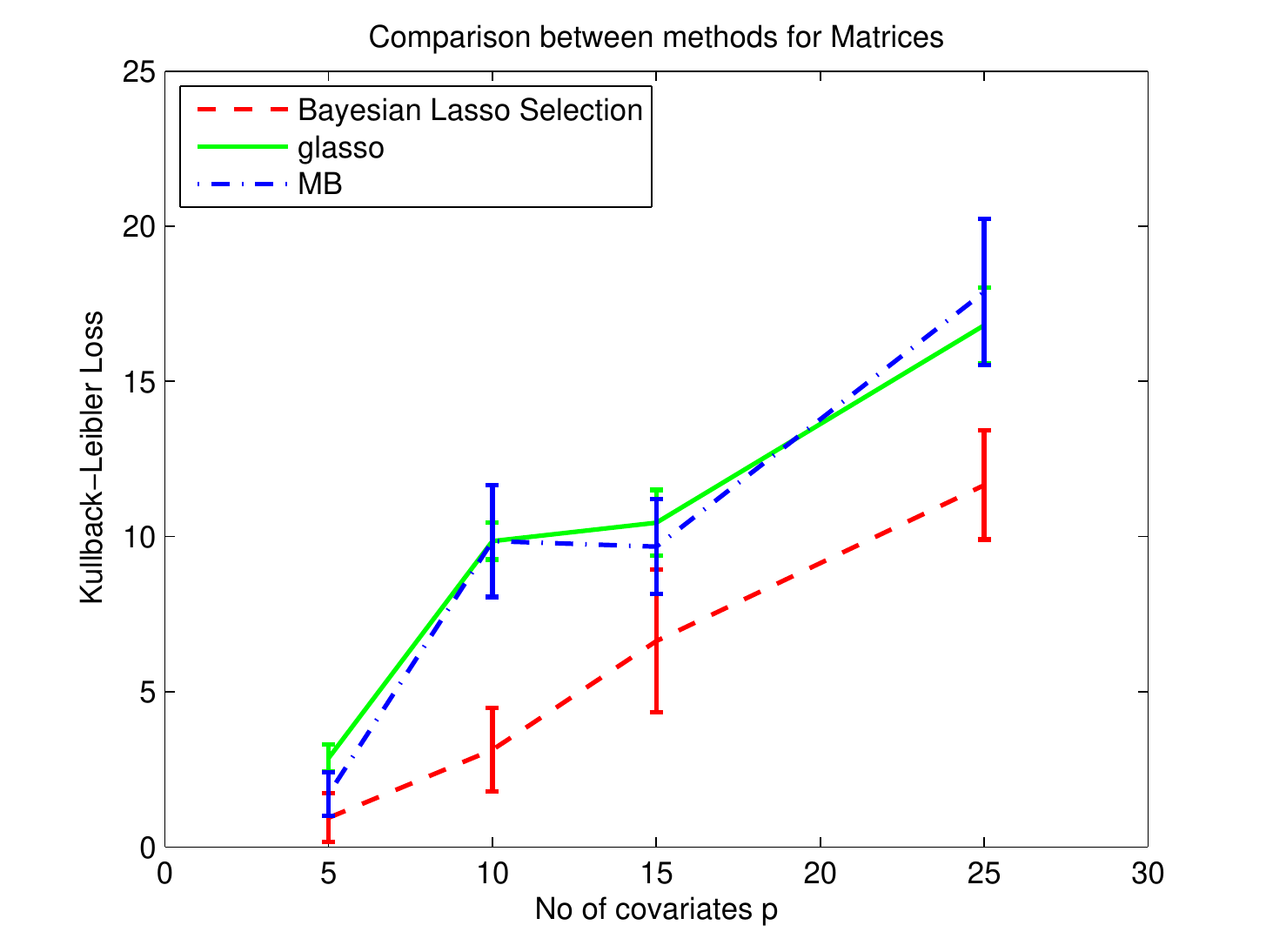}}
\center
\subfigure[Dense   Matrix]{\includegraphics[scale =.5,angle=0]{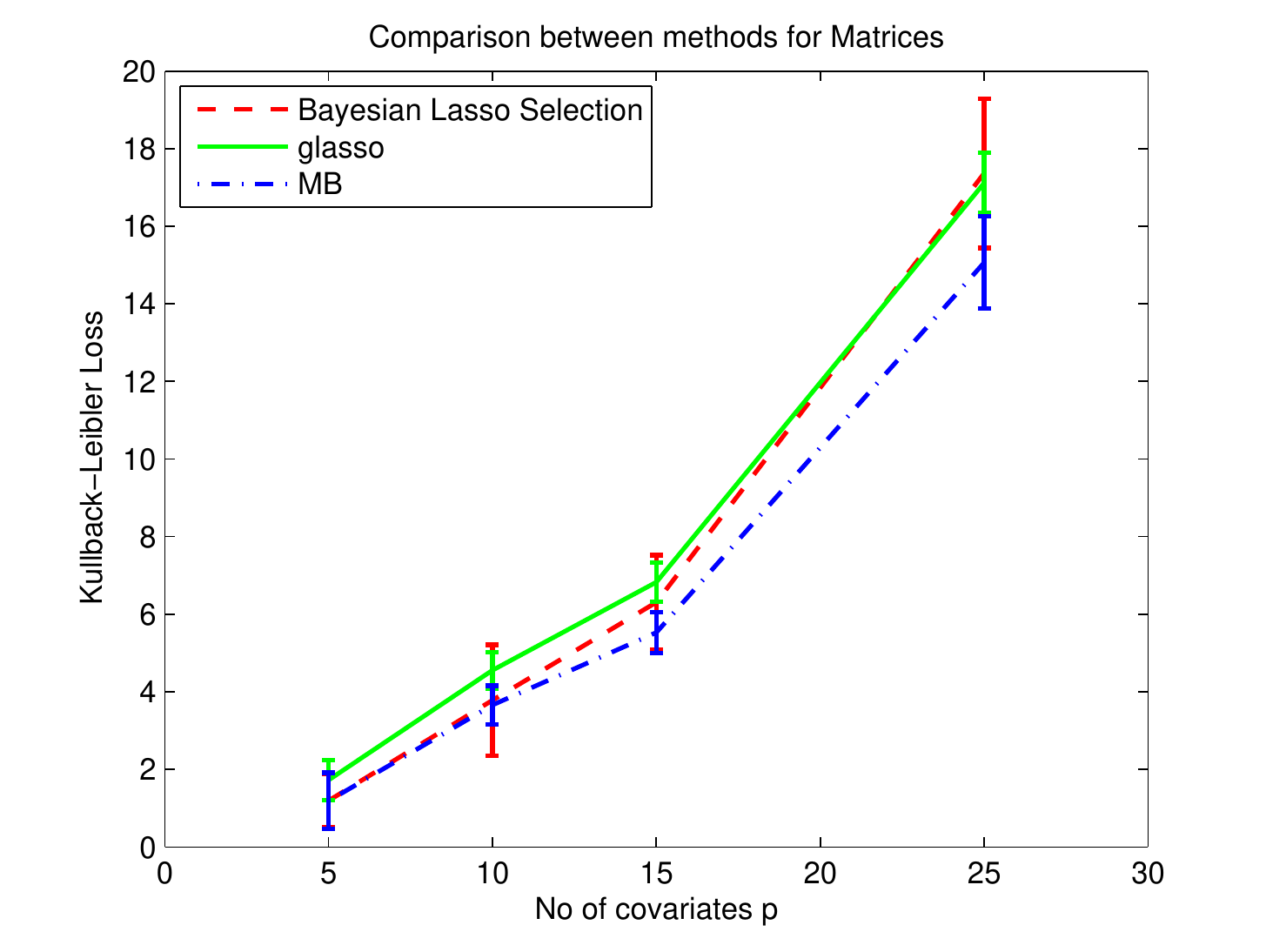}}
\caption{Comparison between 3 methods,  ``glasso", ``MB" and ``Bayesian lasso selection" model in terms of Kullback-Leibler loss for the simulated matrices for different types of structures of the precision matrix;  $p=25$.}
\label{simukl25}
\end{figure}

 The Kullback -Leibler loss is defined as $ \Delta_{KL}(\hat{\bm\Omega},\bm\Omega) = trace(\bm\Omega\hat{\bm\Omega}^{-1}) - log|\bm\Omega\hat{\bm\Omega}^{-1}| - p$. Its ideal value should be zero when $\hat{\bm\Omega} = \bm\Omega$. Figure~\ref{simukl25} shows the means and standard errors for the K-L loss for varying numbers of nodes $ p = \{5,10,15,25\}$  and for sample size $n=25$ averaged over 10 simulations runs. The ``glasso" method and the ``MB" method were performed using $\rho = 0.1$, where $\rho$ is the tuning parameter for the lasso penalty in both methods. This setting yielded the best results for both methods. 

In terms of the K-L loss,  we see that the Bayesian method outperforms the glasso and MB methods for the sparse unstructured and banded diagonal matrices. We believe this is due to the fact that our model allows for local adaptive shrinkage of the partial correlations, whereas the other methods involve global shrinkage based on a single parameter and thus yield sparser solutions. All the methods performed equivalently for the identity matrix, the block diagonal, and the dense structure. For the banded diagonal and sparse structures, which are of greatest interest, the Bayesian method outperformed the frequentist methods.  

\subsection{Comparison with Bayesian Lasso Approaches}
In this section we use simulations to compare the Bayesian Lasso method by \citet{wang2012}. We perform the comparison using the same metrics as presented in \citep{wang2012}. We simulate the following scenarios for comparison.
\begin{enumerate}
\item[1.] Banded diagonal matrix.
\item[2.] Block diagonal matrix.
\item[3.] Sparse unstructured matrix.
\end{enumerate}

The data was simulated using the three matrix structures with  $p=10$ and $n=25$ and are shown in Figure~\ref{simumat}. We compare the Matthews Correlation Coefficient(MCC), sensitivity and specificity of the methods for the three scenarios specified above. The definitions fo the three quantities are as follows:
 Specificity =$\frac{TN}{TN+FP}$, Sensitivity=$\frac{TP}{FN+TP}$ and MCC = $\frac{TN\times TP-FP\times FN} {\sqrt{(TN+FP) (TP+FN)(TP+FP)(TN+FN)    }}$ , where $TN,TP,FN,FP$ are the true negatives, true positives, false negatives and false positives respectively. For the comparison the matlab code from \citet{wang2012} was used with the default settings as provided.
 The results for each of the scenarios are presented in Table 2.

\begin{sidewaystable}[htbp]
  \centering
 \caption{Comparison fo Matthews correlation Coefficient(MCC), Sensitivity and Specificity for the Bglasso method\citep{wang2012} and the Bayesian Lasso selction method presented in this manuscript in three scenarios. The simulated data had 25 samples and 10,15 and 20 covariates.}
    \begin{tabular}{rrrrrrrrrrr}
    Threshold =0.01 & Method & p=10  &       &       & p=15  &       &       & p=20  &       &  \\
    Sparse Matrix &       & MCC   & Sensistivity & Specificity & MCC   & Sensistivity & Specificity & MCC   & Sensistivity & Specificity \\
          & Blasso & 0.23  & 0.93  & 0.27  & 0.23  & 0.88  & 0.43  & 0.17  & 0.83  & 0.42 \\
          & Blasso Selection & 0.3   & 0.53  & 0.76  & 0.27  & 0.41  & 0.87  & 0.22  & 0.59  & 0.73 \\
    Banded Matrix &       & MCC   & Sensistivity & Specificity & MCC   & Sensistivity & Specificity & MCC   & Sensistivity & Specificity \\
          & Blasso & 0.39  & 1     & 0.47  & 0.39  & 1     & 0.58  & 0.37  & 1     & 0.61 \\
          & Blasso Selection & 0.74  & 0.89  & 0.91  & 0.64  & 0.85  & 0.9   & 0.53  & 0.89  & 0.84 \\
    Block Matrix &       & MCC   & Sensistivity & Specificity & MCC   & Sensistivity & Specificity & MCC   & Sensistivity & Specificity \\
          & Blasso & 0.23  & 0.87  & 0.4   & 0.1592 & 0.679 & 0.4722 & 0.3   & 0.91  & 0.55 \\
          & Blasso Selection & 0.67  & 0.63  & 0.97  & 0.55  & 0.59  & 0.86  & 0.37  & 0.86  & 0.68 \\
          &       &       &       &       &       &       &       &       &       &  \\
    Threshold =0.001 & Method & p=10  &       &       & p=15  &       &       & p=20  &       &  \\
    Sparse Matrix &       & MCC   & Sensistivity & Specificity & MCC   & Sensistivity & Specificity & MCC   & Sensistivity & Specificity \\
          & Blasso & 0.15  & 0.33  & 0.5   & 0.06  & 1     & 0.02  & 0.05  & 0.95  & 0.02 \\
          & Blasso Selection & 0.3   & 0.53  & 0.76  & 0.27  & 0.41  & 0.87  & 0.22  & 0.59  & 0.73 \\
    Banded Matrix &       & MCC   & Sensistivity & Specificity & MCC   & Sensistivity & Specificity & MCC   & Sensistivity & Specificity \\
          & Blasso & 0.57  & 1     & 0.7   & NAN   & 1     & 0     & 0.04  & 1     & 0.01 \\
          & Blasso Selection & 0.74  & 0.89  & 0.91  & 0.64  & 0.85  & 0.9   & 0.53  & 0.89  & 0.84 \\
    Block Matrix &       & MCC   & Sensistivity & Specificity & MCC   & Sensistivity & Specificity & MCC   & Sensistivity & Specificity \\
          & Blasso & 0.05  & 0.5   & 0.57  & 0.13  & 0.93  & 0.013 & 0.09  & 0.99  & 0.05 \\
          & Blasso Selection & 0.67  & 0.63  & 0.97  & 0.55  & 0.59  & 0.86  & 0.37  & 0.86  & 0.68 \\
    \end{tabular}%
  \label{tab:addlabel}%
\end{sidewaystable}%

From the above results the proposed Bayesian Lasso selection method outperforms the BGlasso method of \citet{wang2012} in the three scenarios specified above.

\section{Mixtures of Gaussian graphical models}

\subsection{Introduction}
A strength of the proposed method is that it can be employed in a more complex modelling framework in a hierarchical manner. In this section we use the proposed method to develop finite mixture graphical models, where each mixture component is assumed to follow a GGM with an adaptive covariance structure. In Section 6, we extend our methods to infinite mixtures of GGMs using Dirichlet process priors. Our motivation for this model arises from the analysis of a high-throughput genomics data set.  Suppose we have a gene expression data set with $n$ samples and $g$ genes, and are interested in detecting $k$ subtypes of cancer among the $n$ samples.  We assume  a  different network (graph) structure of these $g$ genes for each cancer subtype and have a primary goal of using this information efficiently to  cluster the samples into the correct subtype of cancer. In addition, we wish to identify biologically significant differences among the  networks to explain the variations between the cancer subtypes.  

 

To this end, we extended the Bayesian lasso selection models outlined in the previous sections in a finite mixture set-up. This framework allows for additional flexibility over that of regular GGMs  by allowing a heterogeneous population to be divided into groups that are more homogeneous and  which share similar connectivity or graphs.

\subsection{The hierarchical finite mixture GGM model}
 Let $\bm{Y}_{p\times n} = (\bm{Y}_1,\ldots ,\bm{Y}_n)$ be a $p \times n$ matrix with $n$ samples and $p$ covariates. Here each of the $n$ samples belongs to one of the $K$ hidden groups or strata. Each sample $\bm{Y}_i$ follows a multivariate normal distribution $\bm{N}(\theta_{j},\bm{\Sigma_{j}})$ if it belongs to  the $j$th group. Given a random sample $\bm{Y}_1,\ldots ,\bm{Y}_n$ , we wish to estimate the number of mixtures $k$ as well as the precision matrices $\bm{\Omega_{j}} = \bm{\Sigma_{j}}^{-1},\: j=1,\cdots ,k $. Conditional on the number of mixtures ($K$) 
 we fit a finite mixture model, then vary the number of mixtures and select the optimal number of mixtures using BIC, as explained in Appendix 2.  Alternative ways of model determination are to find the Bayes factor using the MCMC samples \citep{chib2001} or to use infinite mixtures models. The infinite mixtures model has been developed in the next section.
 
 We introduce the latent indicator variable $L_i \in {1,2,\ldots,K}$, which corresponds to  every observation $\bm{Y_i}$ that indicates which component of the mixture is associated with $\bm{Y}_i$, i.e., $L_i=j$ if $\bm{Y}_i$ belongs to the $j^{th}$ group. {\it A priori}, we assume $P(L_i=j)=p_j$ such that $p_1+p_2+\ldots+p_K = 1$. We can then write the likelihood of the data conditional on the latent variables as
$$\bm{Y}_i|L_i=j,\bm\theta,\bm\Omega  \sim  N(\bm\theta_j,\bm\Omega_j^{-1}). $$
The latent indicator variables are allowed {\it a priori} to follow  a multinomial distribution with probabilities $p_1,\ldots,p_K$ as
$$L_i  \sim  \text{Multinomial}(n,[p_1,p_2,\ldots,p_K]),$$
 and the associated class probabilities follow a Dirichlet distribution as
$$ p_1,p_2,\ldots,p_K|\alpha  \sim  \text{Dirichlet}(\alpha_1,\alpha_2,\ldots,\alpha_K).$$
We allow the individual means of each group to follow a Normal distribution  as
$$\bm\theta_j|\bm{B} \sim N(\bm{0},\bm{B}),$$ We assign a common  inverse Wishart prior for covariance matrix $\bm{B}$ across groups  as $\bm{B} \sim IW(\bm{\nu_0},\bm{B_0})$, where $\bm{\nu_0}$ is the shape parameter and $\bm{B_0}$ is the scale matrix. The hierarchical specification of the GGM structure for each group $\bm\Omega_{j}$ parallels the development of the previous section, with each GGM indexed by its own mixture-specific parameters to allow the sparsity to vary within each cluster component.  The hierarchical model for finite mixture GGMs can be summarized as follows:
\begin{align*}
& \bm{Y}_i|L_i=j,\bm\theta,\bm\Omega & \sim &\hspace{1cm} N(\bm\theta_j,\bm\Omega_j^{-1})\\
&L_i & \sim &\hspace{1cm} \text{Multinomial}(n,[p_1,p_2,\ldots,p_K])\\
&p_1,p_2,\ldots,p_K|\alpha &\sim &\hspace{1cm}\text{Dirichlet}(\alpha_1,\alpha_2,\ldots,\alpha_K)\\
& \bm\theta_j|\bm{B} &\sim&\hspace{1cm} N(\bm{0},\bm{B}) \\
&\bm{B} &\sim &\hspace{1cm} IW(\bm{\nu_0},\bm{B_0})\\
& \bm\Omega_j& = &\hspace{1cm}\bm{S}_j(\bm{A_j\odot R_j})\bm{S}_j \\
&A_{j_{(lm)}}|q_{j_{(lm)}}&\sim&\hspace{1cm} \text{Bernoulli}(q_{j_{(lm)}}) ,i<j\\
&\bm{R_j}|\bm{A_j} &\sim&\hspace{1cm} \prod_{l<m}\text{Laplace}(0,\tau_{j_{(lm)}}) I(\bm{C}_j\in \mathbb{C}_p) \\
&\tau_{j_{(lm)}}&\sim& \hspace{1cm}IG(e,f) \\
&S_{j_{(l)}} & \sim & \hspace{1cm} IG(g,h),
\end{align*}
where $i$ denotes the sample, $j$ denotes the mixture component,  $i = 1,2,\ldots,n$ and $j = 1,2,\ldots,K$. In addition,  $A_{j_{(lm)}}$ and $\tau_{j_{(lm)}}$ denote the $lm^{th}$ component of the $\bm{A}_{j}$ and the $\bm{\tau}_{j}$, $l = 1,2,\ldots,p$ , $m= l,\ldots,p$. Posterior inference and conditional distributions are detailed in section 2 of the supplementary material.

\section{Infinite Mixtures of Gaussian Graphical Models}

We have extended our modeling framework to infinite mixtures of Gaussian graphical models in cases where the number of mixture components are unknown. The added advantage of this procedure is the number of mixtures is treated as  a model unknown and will be determined adaptively by data rather than using BIC. Following previous section,  $\bm{Y}_{p\times n} = (\bm{Y}_1,\ldots ,\bm{Y}_n)$ is a $p \times n$ matrix with $n$ samples and $p$ covariates and the likelihood function is
$$\bm{Y}_i|\bm\theta_i,\bm\Omega_i  \sim  N(\bm\theta_i,\bm\Omega_i^{-1}).$$
Let $\gamma_i =(\bm\theta_i,\bm\Omega_i^{-1})$. We propose the Dirichlet process prior \citep{ferguson1973,dey1998, muller2004} on $\bm\gamma=(\gamma_1,\gamma_2,\ldots,\gamma_n)$ which can be written as 
$$\bm\gamma=(\gamma_1,\gamma_2,\ldots,\gamma_n)\sim DP(\alpha,H_{\phi}),$$
where $\alpha$ is the precision parameter $H_{\phi}$ is the base distribution of the Dirichlet process (DP) prior. In a sequence of draws $\gamma_1, \gamma_2, \ldots$ from the Polya urn representation of the Dirichlet process \citep{blackwell1973}, the $n$th sample is either distinct with a small probability $\alpha / ( \alpha + n - 1 )$ or is tied to previous sample with positive probability to form a cluster. Let $\mathbf{\gamma}_{-n} = \{ \gamma_1, \ldots, \gamma_n \} - \{ \gamma_n \}$ and $d_{n - 1}$= number of preexisting clusters of tied samples in $\gamma_{- n}$ at the $n$th draw, then we have 
\begin{equation}
f ( \gamma_n | \mathbf{\gamma}_{- n}, \alpha, \phi ) = \frac{\alpha}{\alpha
+ n - 1} H_{\phi} + \sum_{j = 1}^{d_{n - 1}} \frac{n_j}{\alpha + n - 1}
\delta_{\bar{\gamma}_j}  \label{dpprior},
\end{equation}
where $H_{\phi}$ is the base prior, and the $j$th cluster has $n_j$ tied samples that are commonly expressed by $\bar{\gamma}_j$ subject to $\sum_{j= 1}^{d_{n - 1}} n_j = n - 1$. After $n$ sequential draws from the Polya urn, there are several ties in the sampled values and we denote the set of distinct samples by $\{ \bar{\gamma}_1 , \ldots, \bar{\gamma}_{d_{n}} \}$, where $d_n$ is essentially the number of clusters.

We induce sparsity into the model using the base distribution $H_{\phi}$ which defines the  cluster configuration. 
We allow the individual means of each group to follow a Normal distribution  as
$$\bm\theta_j|\bm{\Omega_j} \sim N(\bm{0},\bm{\Omega_j^{-1}}).$$ The hierarchical specification of the GGM structure for each group $\bm\Omega_{j}$ parallels the development of the previous subsection, with each GGM indexed by its own mixture-specific parameters to allow the sparsity to vary within each cluster component.  Hence, $H_{\phi}=N_{\bm\theta|\Omega}(.) F_{\bm\Omega}$ where $F_{\bm\Omega}$ is the baseline prior for the precision matrix. The hierarchical model for the base prior can be summarized as follows: 

\begin{align*}
H_{\phi} &\propto N_{\bm\theta}(.) F_{\bm\Omega}\\
\bm\Omega&=\bm{S}(\bm{A}\odot \bm{R})\bm{S}\\
F_{\Omega}&\propto F_{A}(.)F_{R}(.)F_{S}(.)\\
\bm{A},\bm{R}|\bm{\tau},\bm{Q}&\sim \prod_{i<j}\text{Laplace}(0,\tau_{ij}) \text{Bernoulli}(q_{ij}) I(\bm{C}\in \mathbb{C}_p) \\
\bm{\tau} &\sim  IG(\nu_{e},\nu_{f})\\
\bm{Q}& \sim Beta(\nu_c,\nu_d)\\
\bm{S} &\sim IG(\nu_{\alpha},\nu_{\beta}).
\end{align*}
The parametrization is similar to the models detailed in the previous sections. The base prior is not in conjugate form so the base prior is not integrable with the likelihood to draw from the posterior using Gibbs sampling framework \citep{escobar1995}. We need to use Metropolis Hastings algorithm to handle the non-conjugacy \citep{neal2000}. Towards this end, we introduce a latent variable, the class indicator for the $i^{th}$ sample, denoted by  $c_i$. We need to update all the $c_i$\rq{s} for each MCMC draw. 
The MCMC state then consists of $\bm c = (c_1,c_2,\ldots,c_n)$. Let $F_Y(.)$ be the data likelihood and $n_c$ be the number of samples in cluster $c$. We use the following algorithm to update the clustering configuration.

\begin{itemize}
\item {\bf New Cluster Creation:} For $i=1,\ldots,n$. If $c_i$ is not a singleton (i.e. $c_i =c_j$ for some $j\neq i$)  , let $c_i^*$ be the new cluster indicator. Draw $\phi_{c_i^*}=[\bm{A,R,\tau,S,\theta,Q}]$ from the base prior $ H_{\phi}$. Probability that a new cluster is created is
$$ p(c_i=c_i^*) = min\{1,\frac{\alpha}{n-1}   \frac{F_Y(Y_i,\phi_{c_i^*})}{F_Y(Y_i,\phi_{c_i})}       \},$$
otherwise, if $c_i$ is a singleton, draw $c_i^*$ from $c_{-i}$ with probability $Pr(c_i^*=c) = n_c/(n-1)$. The the probability of the sample belonging to the cluster $c$ is
$$ p(c_i=c_i^*) = min\{1,\frac{n-1}{\alpha}   \frac{F_Y(Y_i,\phi_{c_i^*})}{F_Y(Y_i,\phi_{c_i})}       \},$$

\item {\bf Existing Clusters:}  For $i=1,\ldots,n$. If $c_i$ is not a singleton, choose a new value for $c_i$  using the following probabilities,
$$Pr(c_i=c) \propto \frac{n_c}{n-1}F_Y(Y_i,\phi_{c}),$$

\item {\bf Cluster Parameters:} Update $\phi_{c}$ for each cluster $c=1,\ldots,d_n$ using $\phi_c|Y_c$ where $Y_c$ are the samples in the cluster $c$. The sampling procedure for updating the cluster parameters is similar to the posterior inference of the  Bayesian lasso selection model.
\end{itemize}

\subsection{Sampling from $H_{\phi}$}

When a new cluster is formed we need to draw the new cluster parameters $\gamma_i =(\bm\theta_i,\bm\Omega_i^{-1})$ from the base prior $H_{\phi}$ which can accomplished as follows:

\begin{itemize}
\item The mean of the cluster $\bm\theta$ can be drawn from the distribution $p(\bm\theta|\bm\Omega)$ which is a multivariate normal distribution. 

\item  As $\bm\Omega=\{\bm{A,R,S,\tau,Q}\}$, we need to draw each of these component to get a draw for $\bm\Omega$.
\item We can sample each of these parameters from the prior distributions specified in a hierarchical manner for the base prior.
\item The probabilities $\bm{Q}$ can be sampled directly from the beta prior.
\item $\bm{\tau}$ can be sampled  directly from the Inverse Gamma prior specified.
\item $\bm{S}$ can be sampled  directly from the Inverse Gamma prior specified.
\item Sampling $\bm{A,R}$ has been done by using griddy Gibbs sampling as described in supplementary material section 1. 
\end{itemize}

\subsection{Real data example}

\begin{figure}[h]
\center{\includegraphics[scale = 0.5]{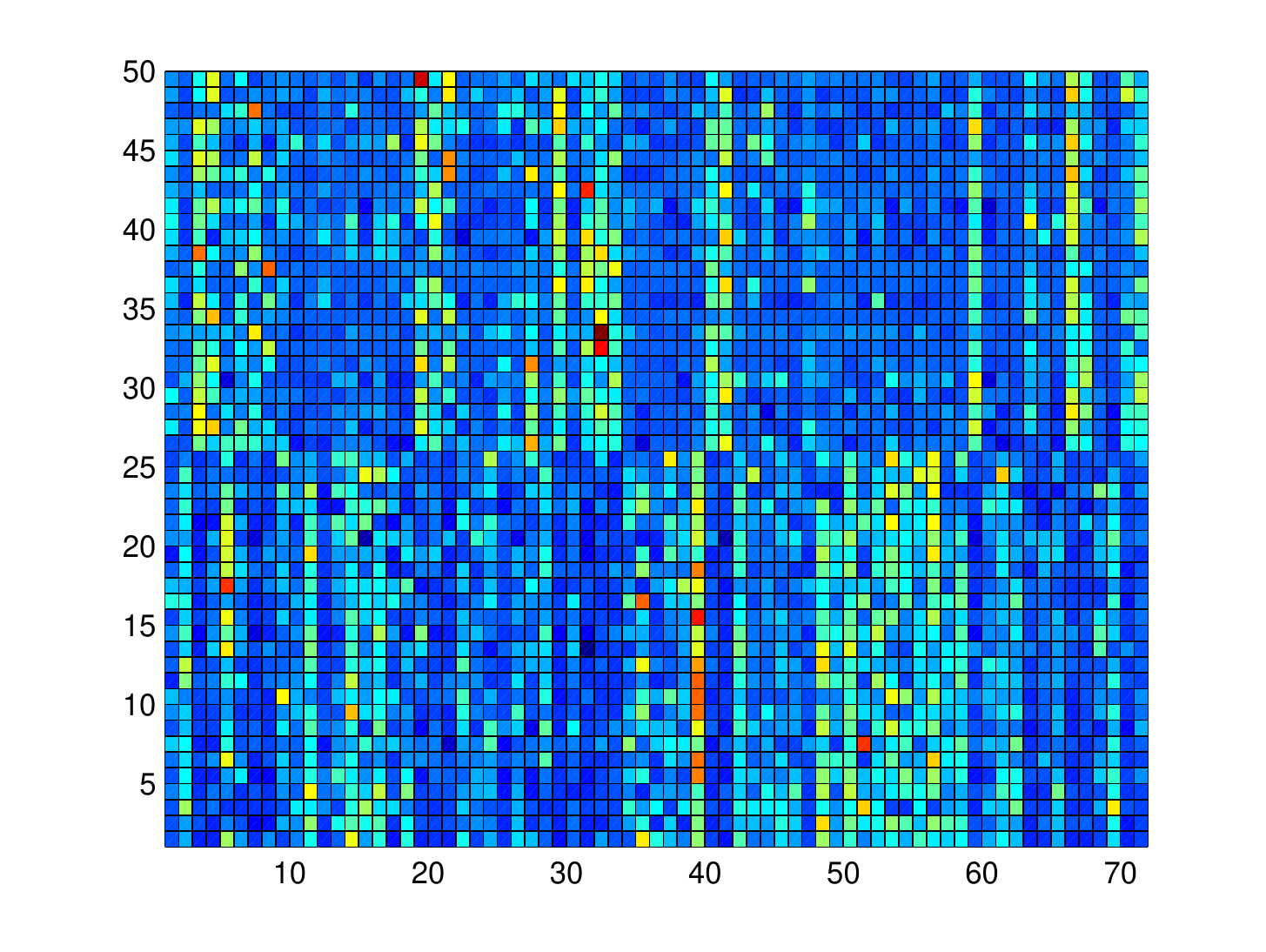}}
\caption{ Heat map of top 50 genes in leukaemia data set.}
\label{luk}
\end{figure}

We used  the leukemia data from \citet{golub99} as a case study to illustrate our mixtures of graphical models. In this study, the authors measured the human gene expression signatures of acute leukaemia. They used supervised learning to predict the type of leukaemia and used unsupervised learning to discover new classes of leukaemia.  The motivation for this work was to improve cancer treatment by distinguishing between subclasses of cancers or tumors. The data are available from \text{http://www.genome.wi.mit.edu/MPR}. The data set includes 6817 genes and 72 patient samples. We selected the 50 most relevant genes, identified using a Bayesian gene selection algorithm \citep{lee02}. The heat map of the top 50 genes in the data set is shown in Figure~\ref{luk} which shows that the expression profiles of these genes form distinct groups of genes that behave concordantly -- hence warranting a more through investigation to explicitly explore the dependence patterns that vary by group.

\begin{figure}[h]
{\includegraphics[width = 14cm]{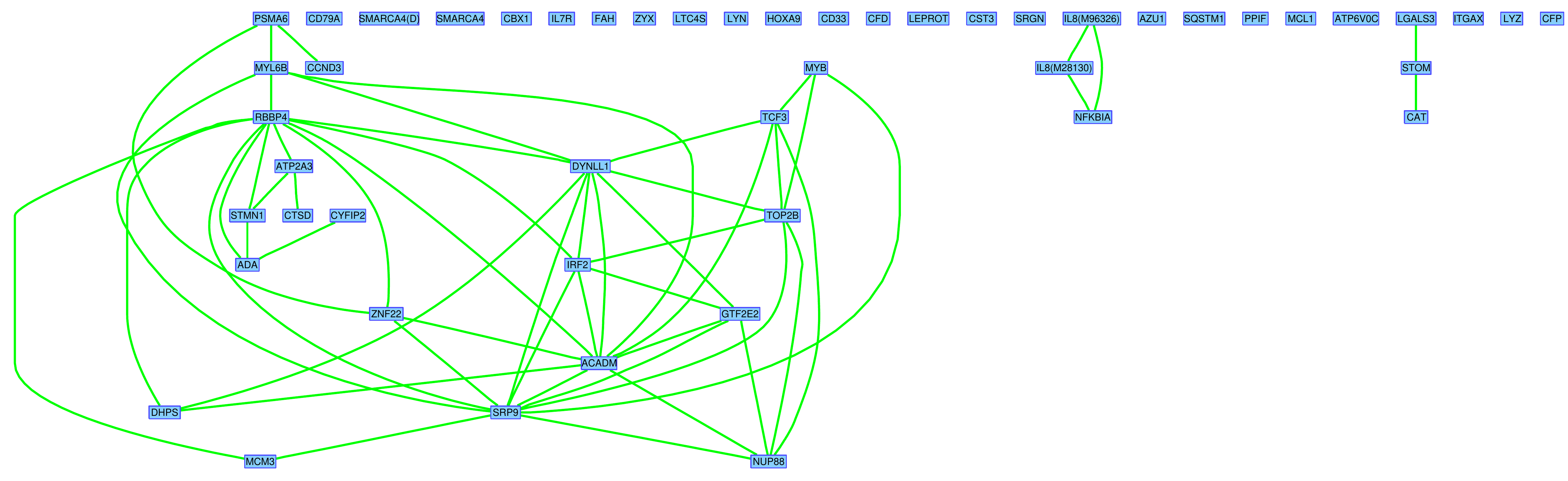}}
\caption{Significant edges for the  genes in the ALL cluster. The red (green) lines between the proteins indicate a negative (positive)  correlation between the proteins.}
\label{all}
\end{figure}

\begin{figure}[h]
{\includegraphics[width = 14cm]{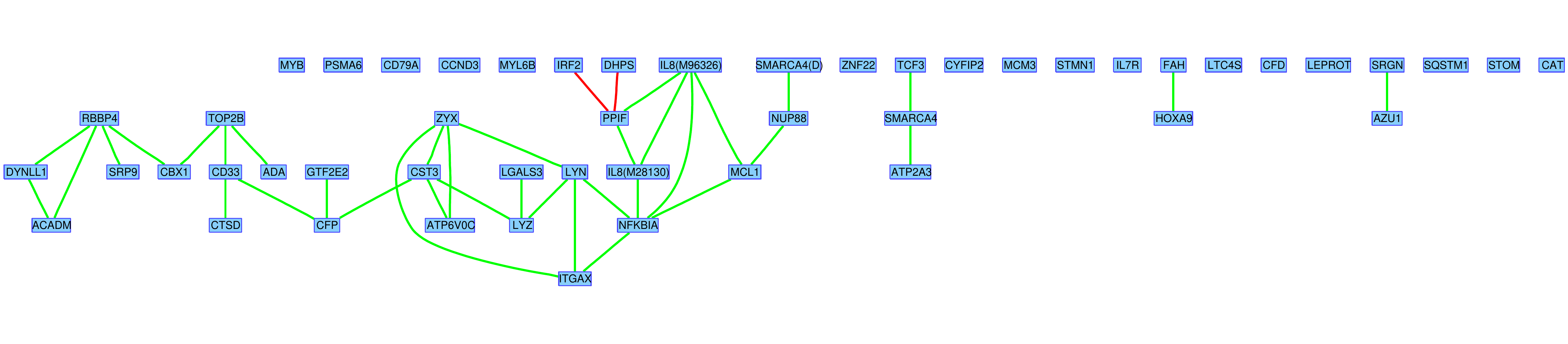}}
\caption{ Significant edges for the  genes in the AML cluster. The red (green) lines between the proteins indicate a negative (positive)  correlation between the proteins.}
\label{aml}
\end{figure}

We fit both finite and infinite mixtures of Gaussian graphical models to this data set. The results are very similar, hence we have produced only the infinte mixtures results.  Using Bayesian lasso selection models and used the methods detailed in Section 3 to find the top graphs for the data. We ran the Markov Chain Monte Carlo simulation for 100000 samples and removed the first 20000 samples as burn-in.   We  obtained  two clusters as corresponding best to two subtypes of leukaemia: (1) acute lymphoblastic leukaemia (ALL) and (2) acute myelogenous leukaemia (AML). The respective networks corresponding to the two clusters are shown in  Figure~\ref{all} and Figure~\ref{aml}. As shown in the figures, the networks for these two clusters are quite different, which suggests  possible interactions between genes that differ depending on the subtype of leukaemia. 

We further explored the biological ramifications of our findings
using the  gene annotations also used by \citet{golub99}.  Most of the genes active in the ALL network are inactive in the AML  network and vice versa. It is known that ITGAX and  CD33 play a role in encoding cell surface proteins which are useful in distinguishing lymphoid from myeloid lineage cells. We can see in the networks of the clusters that CD33 and ITGAX are active in the AML network but inactive in the ALL network. The zyxin gene plays a role in in producing an important protein important for cell adhesion. Zyxin is also active in the AML network but not in the ALL network.  In general, the genes most useful in distinguishing AML vs. ALL class prediction are  markers of haematopoietic lineage, which are not necessarily related to cancer pathogenesis. However, many of these genes encode proteins critical for S-phase cell cycle progression  (CCND3, STMN1, and MCM3), chromatin remodelling (RBBP4 and SMARC4), transcription (GTF2E2), and  cell adhesion (zyxin and ITGAX), or are known oncogenes (MYB, TCF3 and HOXA9)\citet{golub99}. The genes encoding proteins for S-phase cell cycle progression (CCND3, STMN1, and MCM3) were all found to be active in the ALL network but inactive in the AML network. This suggests a connection of ALL with  the S-phase cell cycle. Genes responsible for chromatin remodelling and transcriptional factors were present in both networks, indicating they are common to both types of cancer. This information can be used to discover a common drug for both types of leukaemia.  Among the oncogenes, MYB was related to the ALL network, whereas TCF3 and HOXA9 were related to the AML network.  HOXA9 overexpression is responsible transformation in myeloid cells and causes leukaemia in animal models. A general role for the HOXA9 expression  in predicting AML outcomes has been suggested by \citet{golub99}. We also confirmed that HOXA9 is active in the AML network, but not in the ALL network.

\subsection{Simulations}

We performed a posterior predictive simulation study to evaluate the operating characteristics of our methodology for mixtures of graphical models. We simulated data from our fitted model of the leukaemia data set using the estimated precision matrices for  the two groups, ALL and AML. The  simulation was conducted as follows. Let $(\bm{\hat\mu_j},\bm{\hat\Omega_j^{-1}})$ denote the estimates of the mean and  precision matrices corresponding to the ALL ($j=1$) and AML ($j=2$) groups, respectively, as obtained in the previous section. We generated data under the convolution of the following multivariate normal likelihood, 
$$\bm{Y}_j \sim N(\bm{\hat\mu_j} , \bm{\hat\Omega_j^{-1}}),$$
with 100 samples and 50 covariates. 

\begin{figure}[h!]
\subfigure[ALL]{\includegraphics[scale =.7,angle=0]{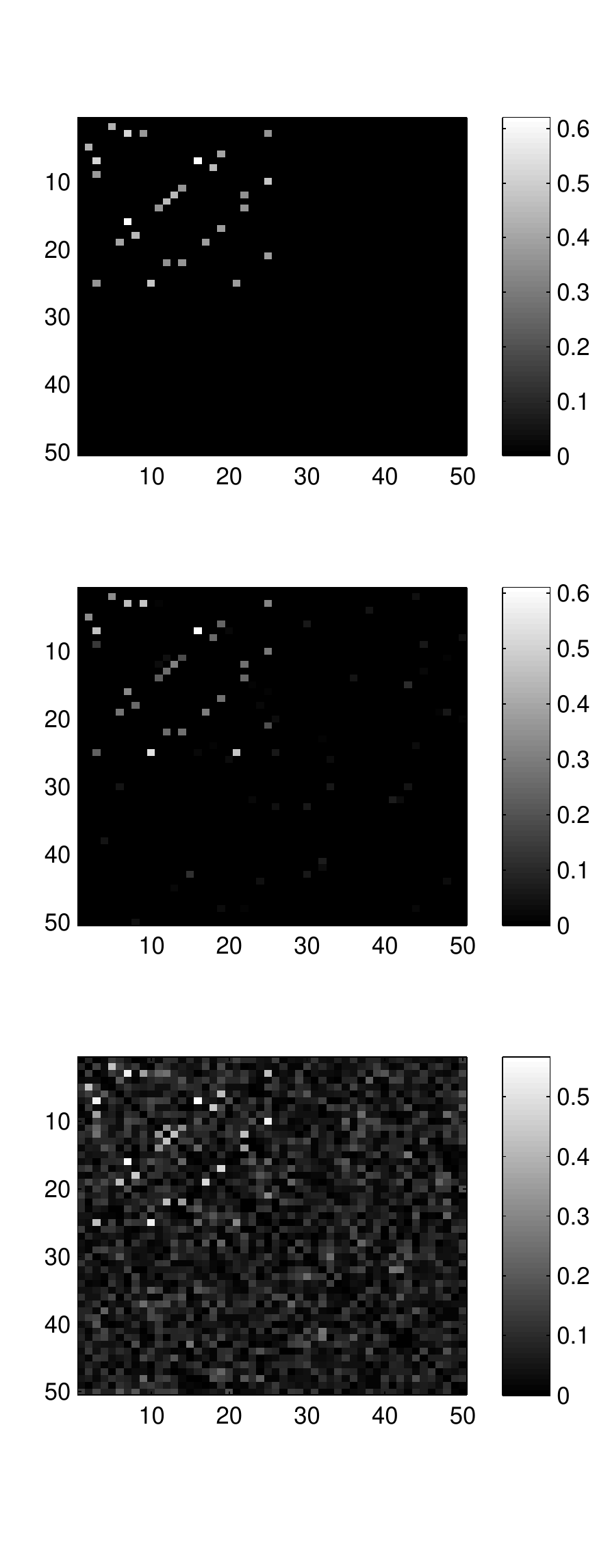}}
\subfigure[AML]{\includegraphics[scale =.7,angle=0]{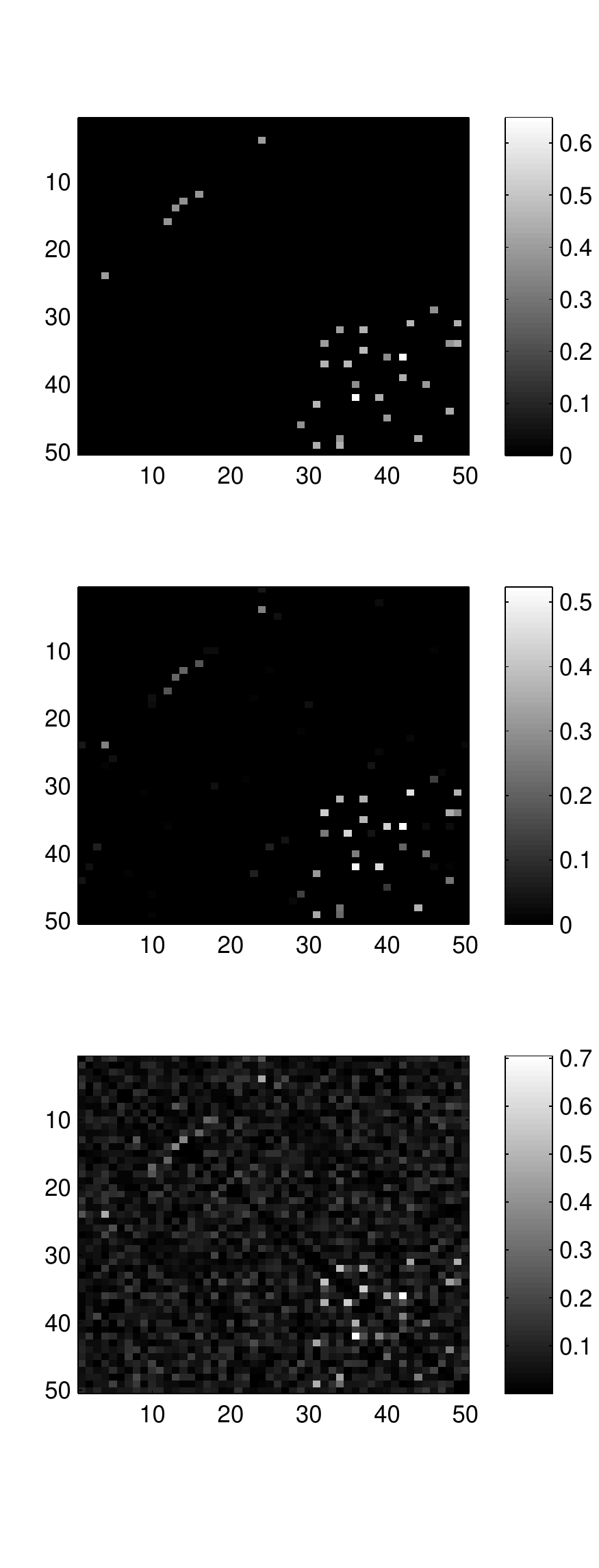}}\\
\caption{Simulation study (p=50). The true and estimated
precision matrices for two subtypes of leukaemia: (a) ALL  and (b) AML. The top row of images shows the true data generating precision matrix; the middle row shows the estimated precision matrix using our adaptive Bayesian model; and the bottom row shows the estimated precision matrix using a non-adaptive fit. Note that the absolute values of the partial correlations are plotted in the above figures without the diagonal. The colorbars are shown to the right of each image. }
\label{gmix}
\end{figure}

We (re-)fitted our models to the simulated data and compared the estimates of the covariance matrices obtained from a non-adaptive finite mixture model (MCLUST) of \citet{fraley07}. We used the ``VVV" setting, which implies the use of an unconstrained  covariance estimation method in their procedure. We completed 100000 runs of the Markov Chain Monte Carlo  simulation and removed the first 10000 runs as burn-in. The true and corresponding estimates of the precision matrices using the two methods  are shown in Figure~\ref{gmix}, where the absolute values of the precision matrix excluding the diagonal are plotted. 

As shown in the figure, fitting our adaptive model to the data (middle row of images)  yields estimates that are closer (sparser) to the true data generating precision matrices, whereas fitting the non-adaptive model to the data   (bottom row of images) yields  noisier estimates, with less local shrinkage of the off-diagonal elements. In addition to a visual inspection, we compared the performance of both methods using  the K-L distance, as in Section \ref{sec:sims}. 
The corresponding estimates of the K-L distances were 3.5592 and 7.2210 for the  adaptive and non-adaptive model fits, respectively. For the AML cluster we obtained respective K-L distances of 4.0836 and 7.7881 for the two methods.  
In addition, we also compared the false positive and false negative rates for finding true edges using each method. It should be noted that the purpose of the MCLUST approach is not covariance selection, hence we imposed selection on the elements of the estimated precision matrix by thresholding the coefficients to zero if they were less than a defined constant. We chose a fairly generous thresholding constant so that the false negatives and false positives were minimized. We applied the thresholding constant of 0.15 to  the coefficients of the precision matrices that were estimated for the two clusters. For the AML cluster, we found false positive rates of (0.0049, 0.0645) and false negative rates of (0.0106, 0) for our adaptive model and the MCLUST approach, 
respectively.   For the ALL cluster, we found false positive rates of (0.0041, 0.0661) and false negative rates of (0.0131, 0) for the adaptive and non-adaptive model fits, respectively. In summary, our adaptive method performs substantially better in recovering the true sparse precision matrix compared to the simple (non-adaptive) clustering approaches.  

\begin{figure}[h!]
\subfigure[ALL]{\includegraphics[scale =.7,angle=0]{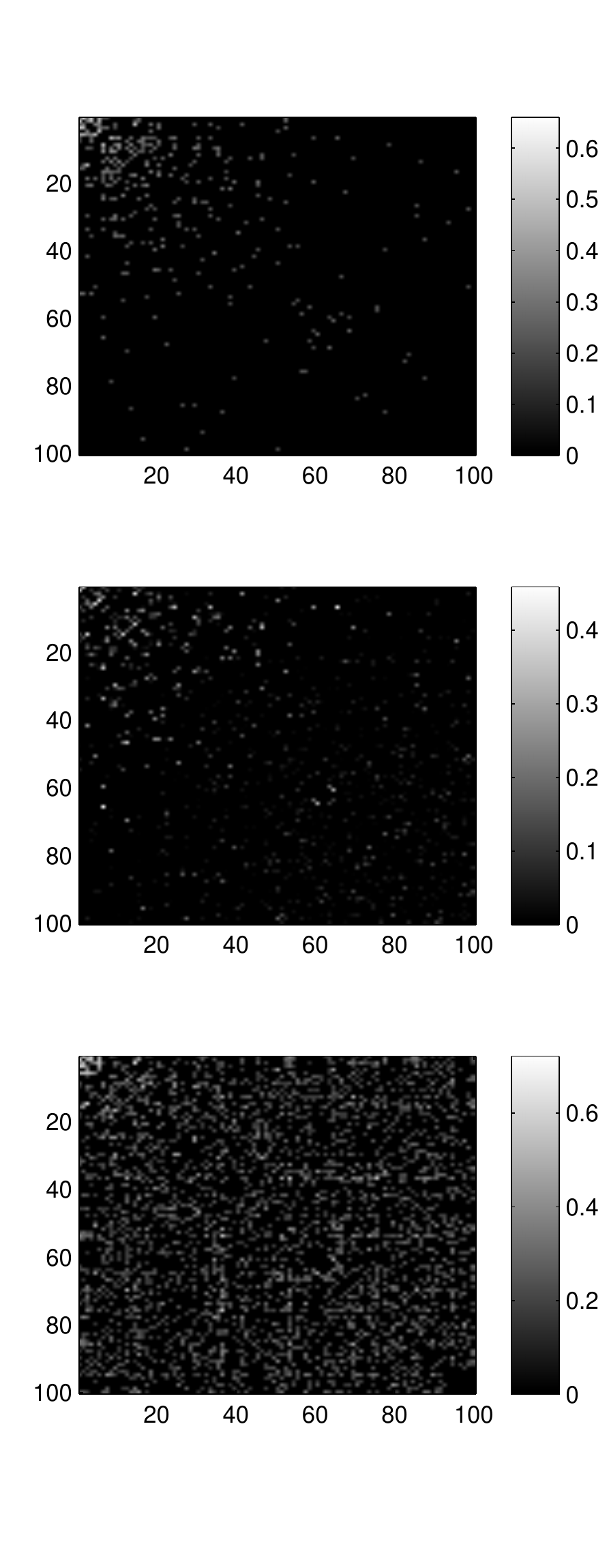}}
\subfigure[AML]{\includegraphics[scale =.7,angle=0]{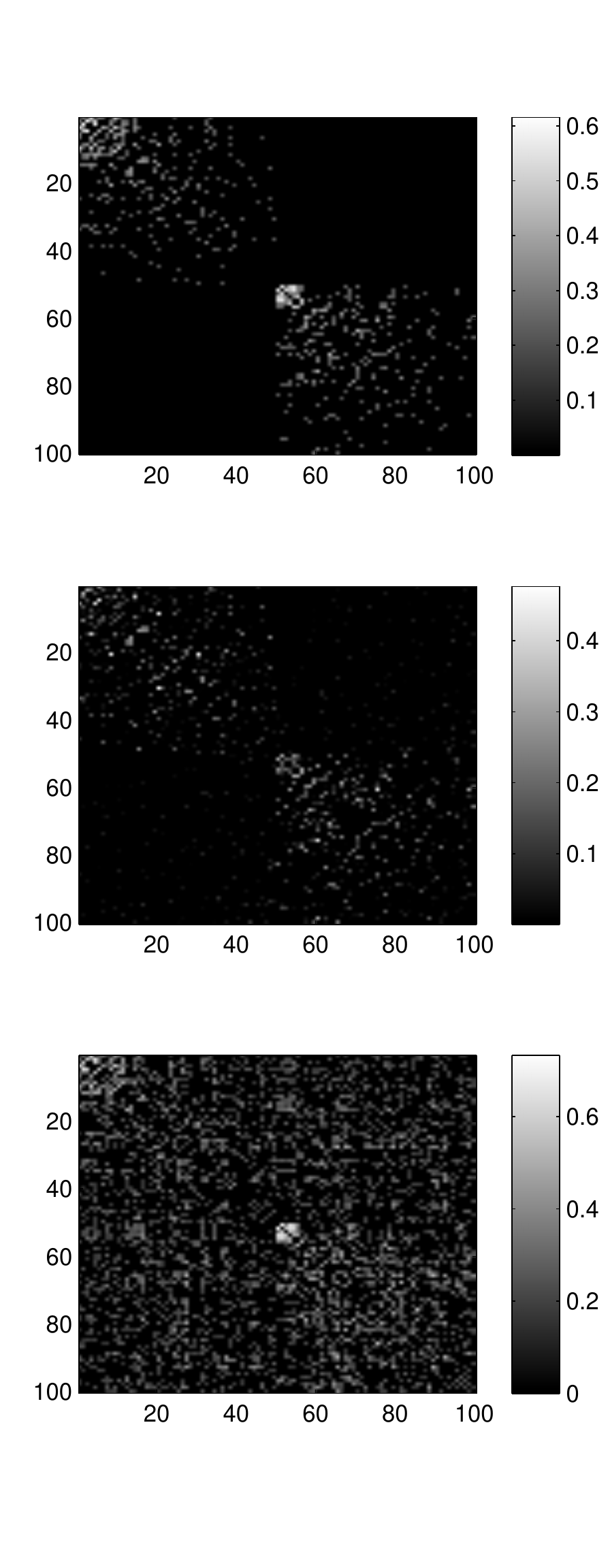}}\\
\caption{Simulation study(p=100). True and estimated
precision matrices for two subtypes of leukaemia: (a) ALL  and (b) AML. The top row of images shows the true data generating precision matrix; the middle row shows the estimated precision matrix using our adaptive Bayesian model; and the bottom row shows the estimated precision matrix  using a non-adaptive fit. Note that the absolute values of the partial correlations are plotted in the above figures without the diagonal. The colorbars are shown to the right of each image.  }
\label{gmix1}
\end{figure}

To explore how the method scales with the number of covariates, we ran another simulation with 100 covariates and 200 samples. The results are plotted in  Figure~\ref{gmix1}. We find a similar pattern of performance from fitting our adaptive model to the data (middle row of images), which  yields estimates that are closer to the true data generating precision matrices. By contrast, fitting the  non-adaptive model to the data  (bottom row of images) yields noisier estimates, with less local shrinkage of the off-diagonal elements. Again we compared the performance of both methods using  the K-L distance and determined that the corresponding estimates were 10.1241 and 25.3378 for the  adaptive and non-adaptive model fits, respectively. For the AML cluster, we obtained  K-L distances of 12.1244 and 27.4851 for the respective methods.  We chose a thresholding constant of 0.15 and applied that to the coefficients of the precision matrices that were estimated for the two clusters. For the AML cluster we found false positive rates of (0.0063, 0.2822) and false negative rates of (0.0222, 0.0081) for our adaptive model and the MCLUST approach, respectively.   For the ALL cluster, we found false positive rates of (0.0044, 0.2497) and false negative rates of (0.101, 0.0372) for the adaptive and non-adaptive fits, respectively. Thus, compared to the non-adaptive approaches, our adaptive method performed substantially better in recovering the true sparse precision matrix. We found that our methods scale reasonably until we reach around 300 covariates but above that level the high computational complexity did not allow for a reasonable computation time. Parallel
computation in cluster machines can be used to speed up the process when the number of covariates extremely high. Alternatively, we plan to explore faster deployments of our algorithm through variational approach or other approximations.



\section{Discussion and conclusions}

In this article we develop a Bayesian framework for adaptive estimation of precision matrices  in Gaussian graphical models. We propose sparse estimators using L1-regularization and use lasso-based selection priors to obtain  sparse and adaptively shrunk estimators of the precision matrix that conduct simultaneous model selection and estimation.   We extend these methods to  mixtures of Gaussian graphical models  for clustered data, with each mixture component assumed to be Gaussian with an adaptive covariance structure. We discuss appropriate posterior simulation schemes for implementing posterior inference in the proposed models, including the evaluation of normalizing constants that are functions of the parameters of interest which result from constraints on the correlation matrix. We compare our methods with several existing methods from the  literature using both real and simulated examples. We found our methods to be very competitive and in some cases to substantially outperform the existing methods.

Our simulations and analysis suggest that it is feasible to implement adaptive GGMs and mixtures of GGMs using Markov Chain Monte Carlo for a reasonable number of variables. Applications to more high-dimensional settings may require more refined sampling algorithms and/or parallelized computations for our method to run in a reasonable time. 

One nice feature of our modelling framework is that it can be generalized to other contexts in a straightforward manner. As opposed to the unsupervised setting we considered, another context would be that of supervised learning or classification using GGMs. Another interesting setting would be to extend our methods for situations in which  the variables are observed over time and our models are used to develop time-dependent sparse dynamic graphs. We leave these tasks for future consideration.

\section*{Acknowledgements}
We would like to acknowledge Ms. LeeAnn Chastain for her help with proofreading the manuscript. This research was supported in part by a cancer prevention fellowship for Rajesh Talluri supported by NIH R25 DA026120 grant from National Institute of Drug Abuse.


\bibliographystyle{asa}
\small{
}

%
%

\end{document}


\baselineskip=14pt

\begin{center}
{\Large{ \bf      Bayesian sparse graphical models and their mixtures using lasso selection priors }}
\end{center}
\baselineskip=12pt

\begin{center}

\end{center}


\pagenumbering{arabic}
\newlength{\gnat}
\setlength{\gnat}{20pt}
\baselineskip=\gnat
\begin{center}
{\Large{\bf Supplementary Material}}
\end{center}

\section{Posterior sampling schemes}
In this model, the $C_{ij}$ is embedded in the likelihood in an awkward manner as well  as the posterior distributions are not of explicit form, hence  we perform the posterior inference using Markov Chain Monte Carlo methods. We derive the full conditionals for all the parameters, and as they are not in  closed form,  we employ the Metropolis-Hastings algorithm to update the model  parameters.  Note that all the full conditionals shown below implicitly assume conditioning on the observed data, $\bm{Y}$.  For simplicity, let $\theta_{ij} = \{\bm{R}_{-ij},\bm{A}_{-ij},q_{ij},\bm{Y}\}$, where $\bm{R_{-ij}}$ and $\bm{A}_{-ij}$ contain all the off-diagonal elements of $\bm{R}$  and $\bm{A}$, respectively, except the $ij^{th}$ element.

The algorithm to perform the Markov Chain Monte Carlo is as follows:

{\bf Sampling of $A_{ij}$ and $R_{ij}$}:\\
The conditional posterior distribution of $R_{ij}$  is 
\begin{eqnarray*}
R_{ij}|A_{ij}=0,\theta_{ij}&\propto & exp\{\frac{1}{\tau_{ij}}|R_{ij}|\})I(\bm{C}\in \mathbb{C}_p)\\
R_{ij}|A_{ij}=1,\theta_{ij}  &\propto& |\bm\Omega|^{n/2} exp\{\frac{-1}{2\sigma^2}tr\{\bm{\Omega Y Y}^T\} - \frac{1}{\tau_{ij}}|R_{ij}|\}I(\bm{C}\in \mathbb{C}_p).
\end{eqnarray*}
where the first equation coincides with the prior distribution on $R_{ij}$ when $A_{ij}=0$. Hence when $A_{ij}=0$, we draw $R_{ij}$ from the prior distribution. Rather than drawing conditionally, we sample $A_{ij}$ and $R_{ij}$ jointly form their conditional posterior distribution \citep{kuo98,della99,della02}. However, any equivalent algorithm such as reversible jump Markov Chain Monte Carlo \citep{green1995} or Carlin and Chib's method (1995) will provide similar results as discussed in \citet{della00} in variable selection context. 
The main difficulty of this algorithm is that  while jointly sampling $[A_{ij},R_{ij}]$ we need to ensure the positive definiteness of $\bm{C}$. The positive definiteness of $\bm{C}$ constrains $C_{ij}|{\rm others}$ to an interval $[u_{ij},v_{ij}]$ \citep{barnard2000} as detailed in Appendix 1.  Once this interval is found, there are several different proposal densities that could be used, such as the uniform or Beta density for performing the Metropolis Hastings algorithm to sample $R_{ij}$'s.

An alternate sampling algorithm that we found worked very well in practice,  instead of using an Metropolis Hastings algorithm (Hastings (1970); Chib and Greenberg (1995)), is to discretize the range interval $[-1,1]$ of $R_{ij}$ into equi-spaced grids. This is an extension of Griddy Gibbs sampling \citep{ritter92} in a bivariate context. In most of our applications, we are mainly interested in the first two digits of a correlation, so that choosing a grid of 100 points is adequate, though extension to finer partitioning or adaptive griding scheme is possible.  Subsequently, we evaluate the joint conditional distribution of $R_{ij}$, $A_{ij}=0$ and $R_{ij}$, $A_{ij}=1$  at these grid values and then make a draw from the normalized bivariate discrete distribution. This way the joint sampling of $A_{ij}$ and $R_{ij}$ is similar to drawing bivariate discrete random variables from a joint probability table, whose one axis contains the values of $R_{ij}$'s (grid values) and the other axis contains the values of $A_{ij}$'s (which is 0 or 1). To emphasize, this conditional distribution becomes 0 for $R_{ij}$, $A_{ij}$ values whose corresponding $C_{ij}$ value lies outside the previously calculated interval $[u_{ij},v_{ij}]$ due to the presence of the indicator function.   The finer details of the algorithm follows:

{\bf Algorithm for the joint sampling of $[A_{ij},R_{ij}]$}:

\begin{itemize}
\item While sampling $[A_{ij},R_{ij}]$ we need to ensure the positive definiteness of $\bm{C}$. 
\item So we first find the range of values $C_{ij}$ can take using the method described in Appendix 1.
\item Suppose that the range of values for $C_{ij}$ to be positive definite is $C_{ij}\in[u_{ij},v_{ij}]$.
\item The joint distribution of $A_{ij},R_{ij}$ is $$(A_{ij},R_{ij}|others)  \propto |\bm\Omega|^{n/2} exp\{\frac{-1}{2\sigma^2}tr\{\bm{\Omega Y Y}^T\} - \frac{1}{\tau_{ij}}|R_{ij}|\}q_{ij}^{A_{ij}}(1 - q_{ij}^{1 - A_{ij}})I(C_{ij}\in[u_{ij},v_{ij}]).$$
\item  Since $R_{ij}$ lies in $[-1,1]$, rather than using the Metropolis Hastings algorithm,  we discretize this interval  in grids and then evaluate the conditional distribution at these grid values.
\item We get a posterior table as follows for $A_{ij}$ and $R_{ij}$ where $P_{A,R}(A_{ij},R_{ij})$ is the value of the unnormalized joint posterior of $[A_{ij},R_{ij}]|others$\\
\end{itemize}

\begin{tabular}{|r|r|r|r|r|r|r|r|}
\hline
 $P_{A,R}(A_{ij},R_{ij})$ &     $R_{ij}=    -1$ &   $R_{ij}=   -0.99 $&               \ldots &    $R_{ij}=   0.99$ &       $R_{ij}=   1$ \\
\hline
      $A_{ij}=  0$ & $P_{A,R}$(0,-1) & $P_{A,R}$(0,-.99) &         \ldots      & $P_{A,R}$(0,.99) &  $P_{A,R}$(0,1) \\
\hline
        $A_{ij}= 1$ & $P_{A,R}$(1,-1) & $P_{A,R}$(1,-.99) &            \ldots      & $P_{A,R}$(1,.99) &  $P_{A,R}$(1,1) \\
\hline
\end{tabular} 
\begin{itemize}
\item The values of the joint posterior outside the valid range of $C_{ij}$ are zero due to the presence of the indicator function.
\item We perform the sampling based on whether the range of  $C_{ij}\in[u_{ij},v_{ij}]$ contains zero or not.
\begin{enumerate}
\item $[u_{ij}<0<v_{ij}]$
\begin{itemize}
\item We jointly draw $A_{ij},R_{ij}$ using the inverse cdf method by sampling one of the configurations of $[A_{ij},R_{ij}]$ from the table after normalizing the probabilities of the configurations.
\item As $A_{ij}=0$ in the first row the likelihood is not affected by $R_{ij}$ and the unnormalized posterior is only affected by the prior on $R_{ij}$. 
\item Hence, when $A_{ij}=0$ the unnormalized posterior density $P_{A,R}$ is equivalent to the prior on $R_{ij}$ which is a Laplace prior.
\end{itemize}
\item $u_{ij},v_{ij}<0$  or $u_{ij},v_{ij}>0$
\begin{itemize}
\item The whole of the first row corresponding to $A_{ij}=0$ has zero entries (i,e, $P_{A,R}(0,R_{ij}) =0$) because $0\notin[u_{ij},v_{ij}]$. . 
\item This condition ensures that $A_{ij}=1$.
\item We  only have to sample from $R_{ij}$ in the second row which corresponds to $A_{ij}=1$.
\item We normalize the unnormalized posterior density of $R_{ij}$ in the second row over the grid on $[-1,1]$ and use the inverse cdf method to sample a value for $R_{ij}$.
\end{itemize}
\end{enumerate}
\item Another equivalent method to jointly sample $[A_{ij},R_{ij}]$ is to draw $A_{ij}$ from its marginal distribution which can be evaluated by summing up the elements of each row of the table. and then draw $R_{ij}|A_{ij}$ by using the inverse cdf method on the  row corresponding to the sampled $A_{ij}$.
\end{itemize}

 {\bf Sampling $\tau_{ij},q_{ij}$}.\\
 The full joint conditional distribution for $\tau_{ij}$ and $q_{ij}$ is
\begin{align*}
\tau_{ij},q_{ij}|A_{ij},R_{ij},\theta_{ij} &\propto K(\tau_{ij},q_{ij}) \frac{1}{\tau_{ij}} exp(\frac{-|A_{ij}R_{ij}|}{\tau_{ij}}) \times \tau_{ij}^{-g-1} exp(-\frac{h}{\tau_{ij}})\\
&\times q_{ij}^{A_{ij}}(1-q_{ij})^{(1-A_{ij})},
\end{align*}
where $K$ is the normalizing constant constrained by the truncation and positive definiteness constraint on $\bm{C} (= \bm{A}\odot\bm{R})$. First, based on $\bm{R}_{-ij}$ we can identify the largest possible interval of $R_{ij}$ that will keep $\bm{C}$ positive definite, say $u_{ij}$ and $v_{ij}$. Specifically,this range of $R_{ij}$ coincides with the range of $C_{ij}$ for $A_{ij}=1$. For $A_{ij}=0$,  $R_{ij}$ has its prior range [-1.1].  Then, we  evaluate $K(\tau_{ij},q_{ij})$ :
\begin{align*}
K^{-1}(\tau_{ij},q_{ij}) &=  \sum_{A_{ij}=\{0,1\}}q_{ij}^{A_{ij}}(1-q_{ij})^{(1-A_{ij})}\int_{-1}^{1} \frac{1}{2\tau_{ij}}exp\{\frac{-|A_{ij}R_{ij}|}{\tau_{ij}}\}I_{[u_{ij},v_{ij}]}(A_{ij}R_{ij})dR_{ij}\\
   & =  \frac{(1-q_{ij})}{2} \frac{(v_{ij}-u_{ij})}{\tau_{ij}}I_{[u_{ij},v_{ij}]}(0) I_{A_{ij}}(0)+ \frac{q_{ij}}{2}W(u_{ij},v_{ij})  I_{[u_{ij},v_{ij}]}(R_{ij}) I_{A_{ij}}(1),
\end{align*}
where $W(u_{ij},v_{ij}) = [   sgn(v_{ij})\{ 1-exp\{\frac{-|v_{ij}|}{\tau_{ij}}\}\}-   sgn(u_{ij})\{ 1-exp\{\frac{-|u_{ij}|}{\tau_{ij}}\}\} ] $and $sgn$ is the sign function
 \begin{equation*}
 sgn(x) = \begin{cases}
-1 & \text{if } x < 0, \\
0 & \text{if } x = 0, \\
1 & \text{if } x > 0. \end{cases}
\end{equation*}

Then we can draw $\tau_{ij}$ and $q_{ij}$ from their  conditional distributions :
\begin{align*}
\tau_{ij}|q_{ij},A_{ij},R_{ij},\bm{Y}&\propto K(\tau_{ij},q_{ij}) \frac{1}{\tau_{ij}} exp(\frac{-|A_{ij}R_{ij}|}{\tau_{ij}}) \times \tau_{ij}^{-g-1} exp(-\frac{h}{\tau_{ij}})\\
q_{ij}|\tau_{ij},A_{ij},R_{ij},\bm{Y} &\propto K(\tau_{ij},q_{ij})q_{ij}^{a_{ij}}(1-q_{ij})^{(1-a_{ij})}q_{ij}^{\alpha-1}(1-q_{ij})^{(\beta-1)}. 
\end{align*}
Both of these conditionals do not have an explicit form, so we need to use the Metropolis Hastings algorithm to draw $\tau_{ij}$ and $q_{ij}$ from their conditionals.\\
 {\bf Sampling $\sigma^2$}.\\
The full conditional distribution of  $\sigma^2$ is in a closed form, so we directly draw from the inverse gamma distribution as
$$k^* = k+np/2,\hspace{5mm} l^*=l+\frac{1}{2} tr(\bm{\Omega Y Y}^T)$$
$$\sigma^2|\bm\Omega,\bm{Y} \sim IG(k^*,l^*).$$

 {\bf Sampling $S_i$}.\\
The full conditional distribution of $S_{i}$ is
\begin{align*}
S_i|\bm{S}_{-i},\bm{Y},\sigma^2 &\propto |\bm{S}(\bm{A}\odot \bm{R})\bm{S}|^{n/2} exp\{ - \frac{1}{2\sigma^2}tr\{\bm{S}(\bm{A}\odot \bm{R})\bm{S}\bm{ Y Y}^T\}\} S_i^{-g-1}exp(\frac{-h}{S_i})\\
& \propto S_i^n exp\{-\frac{1}{2\sigma^2}tr\{\bm{S}(\bm{A}\odot \bm{R})\bm{S}\bm{Y Y}^T\}\} S_i^{-g-1}exp(\frac{-h}{S_i}).
\end{align*}
We use the Metropolis Hastings algorithm to sample $S_i$ from this distribution.

\section{Posterior inference and the conditional distributions for Finite Mixture of Gaussian Graphical Models}
We perform the  posterior inference using Markov Chain Monte Carlo methods; hence  we derive the full conditionals for all the parameters. Not all the full conditionals are  in a closed form; and in those situations we employ the Metropolis Hastings algorithm to simulate those parameters.\\
{\bf Sampling probabilities $p_j$}.\\
We draw the probabilities from a Dirichlet distribution, which can be done by drawing each probability from a gamma distribution with the corresponding Dirichlet parameter and normalizing them so that their sum is equal to 1.
\begin{align*}
&p_1,p_2,\ldots,p_K&|{\rm Others}\propto & \hspace{1cm} \prod_{j=1}^K p_j^{\alpha_j-1}\prod_{j=1}^K p_j^{n_j}\\
&&\sim& \hspace{1cm} Dirichlet(n_1+\alpha_1,n_2+\alpha_2,\ldots,n_K+\alpha_K).
\end{align*}
{\bf Sampling Class Indicators $L_i$}.\\
The full conditional of $L_{i}$ is
\begin{align*}
&P(L_i=j|{\rm Others})&=& \hspace{1cm} \frac{p_j\phi_{Y_i}(\bm\theta_j,\bm\Omega_j^{-1})}{\sum_{j=1}^K p_j\phi_{Y_i}(\bm\theta_j,\bm\Omega_j^{-1})}.
\end{align*}
Each of the class indicators $L_i$ can be drawn from a multinomial distribution with the above probability.\\
{\bf Sampling class means $\bm\theta_j$}.\\
The conditionals for the means of the corresponding mixtures are from a multivariate normal distribution, so we can directly sample them:
\begin{align*}
&\bm\theta_j|{\rm Others}& \propto & \hspace{1cm} N_{\bm\theta_j}(\bm{0},\bm{B})\times \prod_{i=1}^{n_j} N_{\bm{Y}_i}(\bm\theta_j,\bm\Omega_j^{-1})\\
&&\propto & \hspace{1cm} exp(-\frac{1}{2} \bm\theta_j^T \bm{B}^{-1}\bm\theta_j)\times exp(-\frac{1}{2} \sum_{i=1}^{n_j} (\bm\theta_j-\bm{Y}_i)^T \bm\Omega_j(\bm\theta_j-\bm{Y}_i) )\\
&&\propto & \hspace{1cm} exp(-\frac{1}{2} \bm\theta_j^T [n_j\bm\Omega_j+\bm{B}^{-1}]\bm\theta_j-2\bm\theta_j\bm\Omega_j\sum_{i=1}^{n_j}\bm{Y}_i \\
&&&\hspace{2cm} +(\sum_{i=1}^{n_j}\bm{Y}_i)^T\bm\Omega_j[n_j\bm\Omega_j+\bm{B}^{-1}]^{-1} \bm\Omega_j\sum_{i=1}^{n_j}\bm{Y}_i)\\
&&\sim&\hspace{1cm} N_{\bm\theta_j}([n_j\bm\Omega_j+\bm{B}^{-1}]^{-1} \bm\Omega_j\sum_{i=1}^{n_j}\bm{Y}_i,[n_j\bm\Omega_j+\bm{B}^{-1}]^{-1}).
\end{align*}
 The other conditionals are similar to those discussed in the selection model. They are detailed in Appendix 2.
 
\appendix
\section*{Appendix 1}
\subsection*{Validity of $\bm{C}$ as the Gaussian graphical model through parametrization and sampling}

We want to assure that under our parametrization $\bm{C} = \bm{A}\odot \bm{R},$  and sampling scheme, we obtain a proper Gaussian graphical model. It is  clear that there always exist a parametrization like this as we can choose $\bm{A}$ as an identity matrix which will provide the simple, independent Gaussian graphical model with no edge between two vertices. Nonetheless, this  parametrization is not unique as there could be several choices of $\bm A$ and $\bm R$ to produce different Gaussian graphical models. We like to identify the best choice using the data.

Next, we  show that  $\bm{C}$ is a correlation matrix (i.e. symmetric, positive definite matrix with elements lies in the range [-1,1]). As $\bm{A}$ and $\bm{R}$ are symmetric matrices, hence their Hadamard product $\bm{C}$ is a symmetric matrix. Furthermore, range of $\bm{R}$ is $[-1,1]$ and $\bm{A}$ takes the value 0 or 1 so the range of $\bm{C}$ is $[-1,1]$. Therefore,  under this parametrization, $\bm{C}$ is a symmetric matrix taking values in the range $[-1,1]$ and some of its elements could be identically 0. Finally, we have to impose the constraint that $C$ will be a positive definite matrix and we do it through our sampling scheme. 

To satisfy the condition of positive definiteness, we use the constraint $I(\bm{C}\in \mathbb{C}_p)$ in the joint prior distribution of $\bm{A}$ and $\bm{R}$. We perform the MCMC sampling in such a way that the constraint $\bm{C}$ to be positive definite would be satisfied. To implement it, we have drawn $R_{ij},A_{ij}$ sequentially conditioned on other $\bm{R}$ and $\bm{A}$ parameters. Consequently, we need to identify the values of $C_{ij}$ which keep $\bm C$ positive definite given that other parameter values fixed \citep{barnard2000}.

For sake of simplicity, assume we are sampling $c_{12}$ (i.e we are actually sampling $a_{12}$ and $r_{12}$). Now we need to find what values  $c_{12}$  can take for the $\bm{C}$ matrix to be in the space of positive definite correlation matrices assuming all other $c_{ij}$'s known. Lets replace $c_{12}$ by $x$ in the matrix.
$\bm{C} =$$\begin{pmatrix} 
1 &x  &\ldots & c_{1p}\\ 
x &1  &\ldots & c_{2p} \\
\vdots&\vdots&\ddots&\vdots\\
c_{1p} & c_{2p}  &\ldots & 1 \\
\end{pmatrix}$
 The determinant of $\bm{C}$ will be a quadratic function in $x$. which can be written as $f(\bm{C})\equiv f(x) = |\bm{C}| = dx^2+ex+g$, where $d,e$ and $g$ depend on all the other $c_{ij}$'s.
  Let the roots of the equation $f(x)= dx^2+ex+g$ be $u$ and $v$.\\
  \begin{figure}
 \begin{center}\includegraphics[scale = .7]{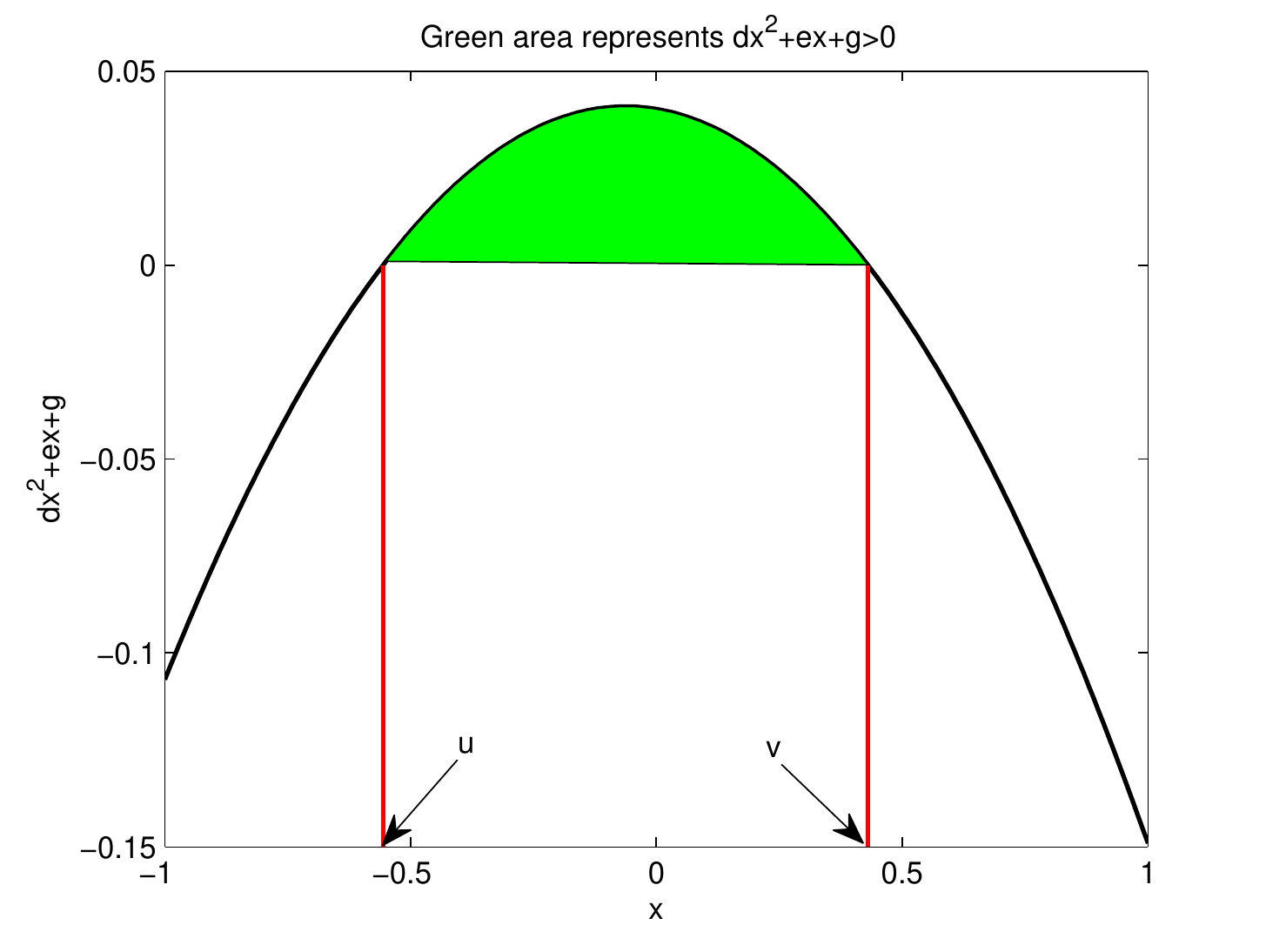}\end{center}
{\bf Fig A1}: This figure illustrates the method used to find the range for $C_{ij}$. The black line represents the quadratic equation of the determinant. $u$ and $v$ represent the roots of the quadratic equation. The range of  values between $u$ and $v$ ensure the positive definiteness of matrix $\bm{C}$.
  \end{figure}
In the above figure the function $f(x)= dx^2+ex+g =0$ is plotted in a black line. The green area represents the valid region for $f(x)>0$. $u,v$ are the roots of the equation where $f(x)=0$. So the range of valid values that $c_{12}$ can take for $\bm{C}$ to belong to the space of positive definite matrices is between $u$ and $v$. The exact expression for the range of $c_{12}$ is:  $$max(-1,u) < c_{12} < min(1,v)$$
  So while sampling $a_{12}$ and $r_{12}$ we have the constraint that $$c_{12}=a_{12}r_{12}\in[max(-1,u),min(1,v)]$$
  Accordingly, this constraint ensures that our sampling scheme generates Gaussian graphical models whose precision matrices are in the space of positive definite matrices.

\section*{Appendix 2}
\subsection*{Conditionals for the finite mixture model}
{\bf Sampling correlation and other parameters related to the precision matrix.}\\
The sampling of all these conditionals is similar to sampling from the previous selection model with slightly different expressions, as we have to sample from each cluster,
\begin{align*}
(R_{j_{(lm)}}|A_{j_{(lm)}},others) & \propto |\bm\Omega_j|^{\frac{n_j}{2}} exp\{-\frac{1}{2}\sum_{i=1}^{n_j}\{ (\bm{Y}_i-\bm\theta_j)^T\bm\Omega_j (\bm{Y}_i-\bm\theta_j)- \frac{1}{\tau_{j_{(lm)}}}|R_{j_{(lm)}}|\}\}I(\bm{C}\in \mathbb{C}_p)
\end{align*}
\begin{align*}
(A_{j_{(lm)}}|R_{j_{(lm)}},others) &  \propto  |\bm\Omega_j|^{\frac{n_j}{2}} exp\{-\frac{1}{2}\sum_{i=1}^{n_j}\{ (\bm{Y}_i-\bm\theta_j)^T\bm\Omega_j (\bm{Y}_i-\bm\theta_j)\}\}q_{j_{(lm)}}^{A_{j_{(lm)}}}(1 - q_{j_{(lm)}}^{1 - A_{j_{(lm)}}}) I(\bm{C}\in \mathbb{C}_p).
\end{align*}
 Here we use approaches similar to those used in the selection model by griding the conditional distribution between $\{u_{j_{(lm)}},v_{j_{(lm)}}\}$ and drawing directly from the conditional.

We draw the $\tau_{j_{(lm)}}$'s  and $q_{j_{(lm)}} $'s using the Metropolis Hastings algorithm. The expression for the normalizing constant  $K(\tau_{j_{(lm)}},q_{j_{(lm)}})$ is similar to the expression given before 

\begin{align*}
\tau_{j_{(lm)}}|q_{j_{(lm)}},A_{j_{(lm)}},R_{j_{(lm)}},\bm{Y}&\propto K(\tau_{j_{(lm)}},q_{j_{(lm)}}) \frac{1}{\tau_{j_{(lm)}}} exp(\frac{-|A_{j_{(lm)}}R_{j_{(lm)}}|}{\tau_{j_{(lm)}}}) \times \tau_{j_{(lm)}}^{-g-1} exp(-\frac{h}{\tau_{j_{(lm)}}})\\
q_{j_{(lm)}}|\tau_{j_{(lm)}},A_{j_{(lm)}},R_{j_{(lm)}},\bm{Y} &\propto K(\tau_{j_{(lm)}},q_{j_{(lm)}})q_{j_{(lm)}}^{A_{j_{(lm)}}}(1-q_{j_{(lm)}})^{(1-A_{j_{(lm)}})}q_{j_{(lm)}}^{\alpha-1}(1-q_{j_{(lm)}})^{(\beta-1)}. 
\end{align*}

Similarly we draw $S_{j_{(l)}}$ using the Metropolis Hastings algorithm from the conditional distribution:
\begin{align*}
S_{j_{(l)}}|\bm{S}_{j_{(-l)}},Y &\propto \prod_{i=1}^{n_j}|\bm{S}_j( \bm{C}_j)\bm{S}_j|^{1/2} exp\{\frac{-1}{2}\{(\bm{Y}_i-\bm\theta_j)^T(\bm{S}_j(\bm{C}_j)\bm{S}_j)(\bm{Y}_i-\bm\theta_j)\}\} \\
 &\times S_{j_{(l)}}^{-g-1}exp(\frac{-h}{S_{j_{(l)}}}).
\end{align*}

\section*{Appendix 3}

\subsection*{Computing BIC values for the graphs}
The Bayesian information criterion (BIC) is widely used for model selection problems.\citep{carlin1995,chib1995} BIC penalizes the complex models in favor of balanced models. BIC can be computed as 
$$ -2\log p(Y|\mathcal{G})+const\approx -2L(Y,\hat\theta)+ m_\mathcal{G}log(n) \equiv BIC,$$
where $p(Y|\mathcal{G})$ is the  likelihood of the data for the model $\mathcal{G}$, $L(Y,\hat\theta)$ is the
maximized log likelihood for the model, $m_{\mathcal{G}}$ is the number of independent parameters to be estimated in the model, and $n$ is the number of samples. Given any two estimated models, $\mathcal{G}_1$ and $\mathcal{G}_2$, the model with the lower value of BIC is the preferred model. The number of parameters to be estimated in the model is considered to be the number of non-zero edges and all the other parameters in the model. In the finite mixture model the number of clusters is not considered an independent parameter for the purpose of computing the BIC. If each model is equally likely {\it a priori}, then $p(Y|\mathcal{G})$ is proportional to the posterior probability that the data conform to the model $\mathcal{G}$.

\section*{Appendix 4}
\subsection*{Implementation of BGlasso code of Wang(2012)}
The code used  to implement the BGlasso is as follows.

\% wangsim\\
burnin  = 1000; nmc = 2000;\\
\% (1) Bayesian Graphical Lasso (Friedman et al 2009 Biostatistics)\\
alambda = 1; blambda = 0.1; \% Prior hyperparameters\\
Y = dataset';\\
S = Y'*Y;\\
Sig = S; C = inv(Sig); \% Initial values \\
(Sigsave,Csave,lambdasave) = BayesGLassoColumnwise(S,n,Sig,C,alambda,blambda,burnin,nmc);\\
for runs = 1:size(Csave,3)\\
Psave(:,:,runs) = corrcov((Csave(:,:,runs)));\\
end\\
precwang = mean(Psave,3);\\
precwang = precwang.*(abs(precwang)$>$.001);\\

The burn in was used as 1000 iterations. The actual MCMC used were 2000 iterations from the given example.The prior hyper parameters were the default as used in the example.We converted the covariance into correlation by standardizing the precision matrix. The mean of the precision matrix was taken as the final estimate of the precision matrix. The precision matrix was thresholded using a threshold of 0.001 as presented in the simulation example in Wand(2012).